\documentclass{article}

\usepackage{iclr2024_conference,times}
\iclrfinalcopy

\usepackage[utf8]{inputenc} 
\usepackage[T1]{fontenc}    
\usepackage{hyperref}       
\usepackage{url}            
\usepackage{booktabs}       
\usepackage{amsfonts}       
\usepackage{nicefrac}       
\usepackage{microtype}      
\usepackage{xcolor}         
\hypersetup{colorlinks=true}
\usepackage[page]{appendix}
\usepackage{caption}
\usepackage{subcaption}
\usepackage{graphicx}
\usepackage{algorithm}
\usepackage{algpseudocode}
\usepackage{fancybox,framed}
\usepackage{multirow}
\usepackage{tabularx}
\usepackage{soul}
\usepackage{tcolorbox}
\usepackage{marvosym}
\usepackage{amsmath}

\newcommand\boldgreen[1]{\textcolor[HTML]{0F9D58}{\textbf{#1}}}

\title{DiarizationLM: Speaker Diarization Post-Processing with Large Language Models}

\author{%
  Quan Wang$^\star$ \quad Yiling Huang$^\star$ \quad Guanlong Zhao$^\star$ \\
  \bf Evan Clark \quad Wei Xia \quad Hank Liao \\
  Google LLC, USA \qquad $^\star$Equal contribution\\
  \Letter $\;$ \texttt{quanw@google.com} \\
}

\fancypagestyle{firstpage}{
  \lhead{\begin{picture}(0,0)\put(0,-3){\includegraphics[width=0.16\linewidth]{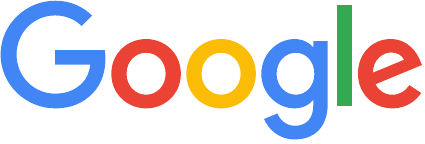}}\end{picture}}
}

\begin{document}

\maketitle
\thispagestyle{firstpage}
\begin{abstract}
In this paper, we introduce DiarizationLM, a framework to leverage large language models (LLM) to post-process the outputs from a speaker diarization system. Various goals can be achieved with the proposed framework, such as improving the readability of the diarized transcript, or reducing the word diarization error rate (WDER). In this framework, the outputs of the automatic speech recognition (ASR) and speaker diarization systems are represented as a compact textual format, which is included in the prompt to an optionally finetuned LLM. The outputs of the LLM can be used as the refined diarization results with the desired enhancement. As a post-processing step, this framework can be easily applied to any off-the-shelf ASR and speaker diarization systems without retraining existing components. Our experiments show that a finetuned PaLM 2-S model can reduce the WDER by rel. $55.5\%$ on the Fisher telephone conversation dataset, and rel. $44.9\%$ on the Callhome English dataset.

\textbf{Code:} {\scriptsize \url{https://github.com/google/speaker-id/tree/master/DiarizationLM}}

\textbf{Model:} {\scriptsize \url{https://huggingface.co/diarizers-community/DiarizationLM-Gemma-4-E4B-v1}}

\textbf{Demo:} {\scriptsize \url{https://huggingface.co/spaces/diarizers-community/DiarizationLM-GGUF}}
\end{abstract}

{
  \hypersetup{linkcolor=black}
  \tableofcontents
}

\section{Introduction}
\label{sec:intro}
Speaker  diarization  is  the  task  of  partitioning speech into homogeneous  segments  according  to  speaker  identities, answering the question ``who spoken when''~\cite{park2021review,zhang2022odysseytutorial}. Typical speaker diarization systems can be roughly categorized into two groups: modularized systems and end-to-end systems. A modularized speaker diarization system usually consists of multiple separately trained components including voice activity detection (VAD)~\cite{zazo2016feature,medennikov2020target,ding2019personal,ding2022personal}, speaker turn detection~\cite{xia2022turn,zhao2022augmenting}, speaker encoder~\cite{ge2e,li2017deep,snyder2018x}, and a clustering algorithm, which can be either unsupervised~\cite{wang2017speaker,garcia2017speaker,dimitriadis2017developing,park2019auto,park2021multi,wang2022highly} or supervised~\cite{zhang2019fully,li2019discriminative}. End-to-end systems , on the other hand, directly optimize the entire system on diarization errors by introducing a permutation invariant loss function~\cite{fujita2019end,ueda2022eend,e2ediarizationpatent,horiguchi2020end}.

In many real world applications such as meeting summarization, call center analysis, mobile recorder apps~\cite{speakerlabelsblog}, and video captioning, knowing ``who spoke when'' is not sufficient. Speaker labels are more interpretable and meaningful when they are associated with speech transcripts. Various solutions have been proposed to directly address the problem of ``who spoke what'', including jointly training speech recognition and speaker diarization~\cite{shafey2019joint}, speaker-attributed automatic speech recognition (SA-ASR)~\cite{kanda2020joint,kanda2021minimum,kanda2021end,kanda2022streaming}, target speaker automatic speech recognition (TS-ASR)~\cite{zmolikova2017speaker,delcroix2018single,delcroix2019end,kanda2019auxiliary} and word-level end-to-end neural speaker diarization~\cite{huang24d_interspeech}.

In practice, however, most production speech systems still consist of separately trained ASR models and speaker diarization models, with various considerations including:
\begin{enumerate}
    \item \emph{Modularized development and deployment:} ASR and speaker diarization systems are usually trained on different datasets, and potentially using different modeling framework, by different research teams. 
    \item \emph{Potential quality regression on ASR:} ASR has many more use cases than speaker diarization. Joint modeling of ASR and speaker diarization usually has worse Word Error Rates (WER) than ASR-only models, thus is not acceptable in many applications.
    \item \emph{Flexibility:} Combining separately trained ASR models and speaker diarization models is a very flexible solution. As long as the ASR model provides word timing information, it can be combined with almost any speaker diarization model, either unsupervised or supervised, either modularized or end-to-end trained. 
\end{enumerate}

\begin{figure}
	\centering
	\begin{subfigure}{\textwidth}
		\includegraphics[width=0.94\linewidth]{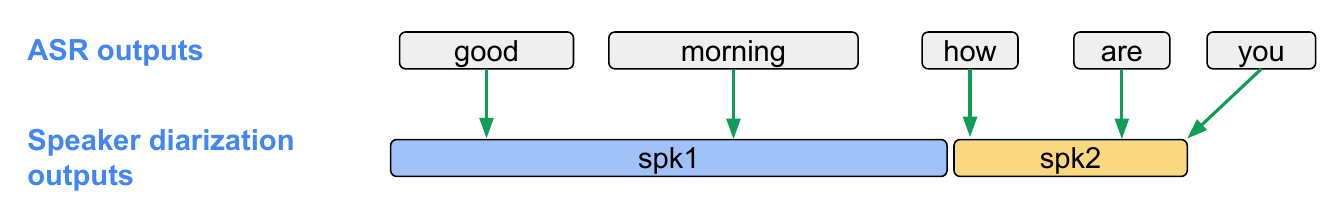}
		\caption{}
		\label{fig:good_orchestration}
	\end{subfigure}
	\newline
	\begin{subfigure}{\textwidth}
		\includegraphics[width=\linewidth]{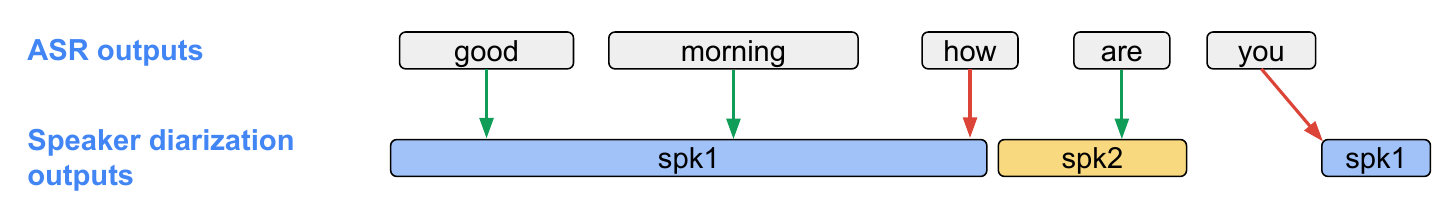}
		\caption{}
		\label{fig:bad_orchestration}
	\end{subfigure}
	\caption{The orchestration module associates each word from the ASR transcript with a speaker label from the speaker diarization outputs. (a) In this example, all words are associated with the correct speaker labels (\textcolor[HTML]{0f9d58}{green} arrows). The words ``good'', ``morning'', and ``are'' and ``you'' are associated with the only speaker label that overlap with them. The word ``how'' overlaps with both spk1 and spk2, but has bigger overlaps with spk2, thus is associated with spk2. The word ``you'' does not overlap with any speaker, but is closest to spk2, thus is associated with spk2. (b) In this example, two words are associated with wrong speaker labels (\textcolor[HTML]{db4437}{red} arrows) due to inconsistent timing information from the two systems. The word ``how'' is mistakenly associated with spk1, since spk1 has more overlap with this word than spk2. The word ``you'' is mistakenly associated with spk1, since spk1 is closer to this word than spk2.}
	\label{fig:orchestration_examples}
\end{figure}

We refer to the combination of ASR transcripts and speaker diarization results as an \emph{orchestration module} (in some other work~\cite{paturi2023lexical}, this process is called ``reconciliation''). In this module, each word from the ASR transcript is associated with a speaker label. A typical orchestration algorithm works as follows: (1) If the word segment overlaps with at least one speaker segment, then this word is associated with the speaker that has the biggest temporal overlap with this word; (2) otherwise if this word segment does not overlap with any speaker segment, then it is associated with the speaker that has the smallest temporal distance to this word based on the segment boundaries. This orchestration algorithm is illustrated in Fig.~\ref{fig:good_orchestration}.

However, since ASR and speaker diarization are separately trained with usually different training datasets and modeling approaches, the timing information from these two systems can be inconsistent, resulting in word diarization errors, as demonstrated with the example in Fig.~\ref{fig:bad_orchestration}. Specifically, modern ASR models are usually trained end-to-end without using the ground truth timing information, and the word timing is inferred from the probability lattice of the decoder, which could be inaccurate.

In many cases, such errors can usually be fixed by leveraging semantic information from the ASR transcripts. Take Fig.~\ref{fig:orchestration_examples} as an example, simply by looking at the textual transcript ``good morning how are you'', if we know it consists of two speakers, we can easily tell which word comes from which speaker confidently without using any acoustic speaker diarization system. In practice, diarization errors can be much more complicated than the simple example in Fig.~\ref{fig:orchestration_examples}. To handle such cases,  we propose DiarizationLM, a framework to post-process the orchestrated ASR and speaker diarization outputs with a large language model (LLM). While the experiments performed in this paper mainly focus on reducing word diarization errors using LLM, we also show examples how LLM can be used for other purposes such as autofilling speaker names, autofilling speaker roles, or even completely replacing the orchestration module in Appendix~\ref{sec:other_tasks}.

\section{Motivating example}
\label{sec:motivation_example}

Here is a small example where we construct a prompt (in \textcolor{blue}{blue}) which consists of a brief instruction, and the diarization transcript with several errors. The prompt is sent to Google Bard~\cite{bard} via its Web UI\footnote{We used an internal version of Bard that is based on a larger model and supports more tokens than the public version.}. We can see that the completion generated by LLM (in \textcolor{orange}{orange}) correctly moved the error words to the right speakers. This example can be easily reproduced with many off-the-shelf commercial or open source instruction-finetuned LLMs, such as Google Bard~\cite{bard} or OpenAI ChatGPT~\cite{chatgpt}.

\begin{framed}
\textbf{Prompt:}

\textcolor{blue}{
In the speaker diarization transcript below, some words are potentially misplaced. Please correct those words and move them to the right speaker. Directly show the corrected transcript without explaining what changes were made or why you made those changes. \\
 \\
<spk:1> Good morning Patrick, how \\
<spk:2> are you? Good, good. How are you Tom? Pretty \\
<spk:1> good. Going to work? \\
<spk:2> Yes. Busy day. How are your kids? Do they go  \\
<spk:1> to school? Oh they are too young for that. I sent them to daycare earlier \\
<spk:2> today. Oh yeah I forgot about that.
}

\textbf{Completion:}

\textcolor{orange}{
<spk:1> Good morning Patrick, how are you? \\
<spk:2> Good, good. How are you Tom? \\
<spk:1> Pretty good. Going to work? \\
<spk:2> Yes. Busy day. How are your kids? Do they go to school? \\
<spk:1> Oh, they are too young for that. I sent them to daycare earlier today. \\
<spk:2> Oh yeah, I forgot about that.
}
\end{framed}

Apart from correcting diarization errors, we also found several other diarization capabilities of LLMs, such as autofilling speaker names, autofilling speaker roles, or even completely replacing the orchestration module. Examples of these capabilities are shown in Appendix~\ref{sec:other_tasks}. However, due to lack of relevant training and testing datasets, these capabilities are not thoroughly studied or evaluated in this paper. Here we focus on LLM's capability of reducing diarization errors. 

\section{DiarizationLM}

\subsection{System overview}

\begin{figure}
	\centering
	\includegraphics[width=\linewidth]{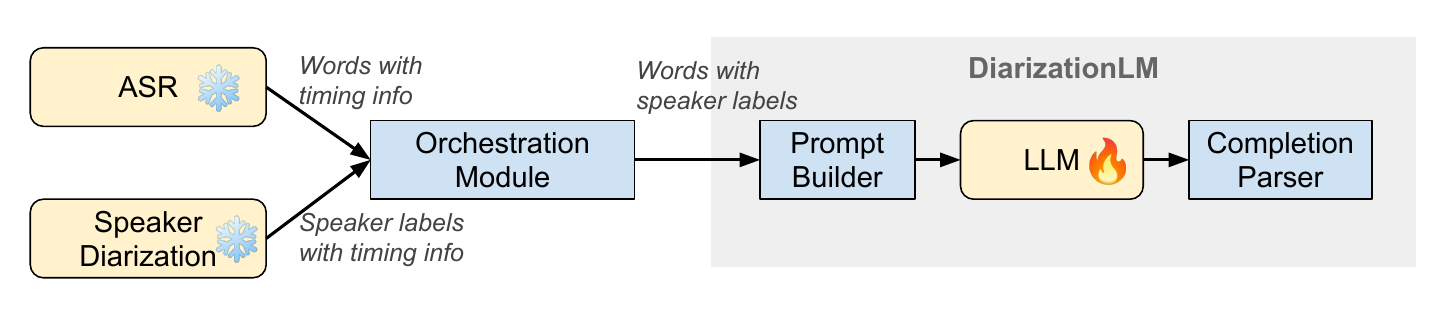}
	\caption{Diagram of the proposed DiarizationLM framework.}
	\label{fig:diagram}
\end{figure}

We illustrate the DiarizationLM framework in Fig.~\ref{fig:diagram}.
In this framework, the ASR and speaker diarization systems are frozen, and their outputs are processed by the orchestration module to associate a speaker label with each recognized word. The orchestrated diarization outputs are processed by a \emph{prompt builder} module, which creates a compact textual representation of the diarized transcript, segment it into shorter versions to fit the LLM input size limit, and apply prompt prefix and suffix. The prompts are then sent to a finetuned LLM, and the completions generated by the LLM will be handled by a \emph{completion parser} module, which truncates undesired outputs from the LLM, combines the completions of multiple segments, and apply a transform (see Section~\ref{sec:tpst}) to preserve the original transcripts of the ASR model.

\subsection{Prompt builder}

The output of the orchestration module is two sequences of equal length: a sequence of words, and a sequence of speaker labels. To fit it into a prompt, we use a compact textual representation, where speaker tokens are only inserted in the beginning of the transcript, or when the speaker has changed. Below is an example:

\begin{tcolorbox}[title=\bf{Example text representation of the prompt}, colback=blue!5!white, colframe=blue!75!black,parbox=false]
\small
\texttt{
\begin{tabular}{l l}
Word sequence: & ["good", "morning", "how", "are", "you"]\\
Speaker sequence: & [1, 1, 2, 2, 2]\\
Text representation: & "<spk:1> good morning <spk:2> how are you"
\end{tabular}
}
\end{tcolorbox}

Since most LLMs have an input length limit, the text representation of an entire utterance may not fit this limit. In such cases, we recursively binary partition the word and speaker sequences in the middle, until all segments fit the the input length limit.

We also apply prefix and suffix to each prompt. The prefix is usually an instruction describing the task for the LLM to perform, and the suffix is a sequence of tokens to indicate the end of the prompt.

\subsection{Completion parser}
Each prompt from the prompt builder will be sent to the finetuned LLM, which will generate a text completion for this prompt. First of all, we need to truncate any undesired outputs from the LLM. For example, during the LLM finetuning, each completion may have a suffix to indicate the end of the completion. Thus the suffix and any text generated after the suffix should be truncated from the original completion.

After the truncation, we need to convert the text representation of the completion back to the word sequence and the speaker sequence format. If the text representation does not start with a speaker token, we either use the last speaker from the previous segment, or just use speaker 1 if it is the first segment.

Next, we concatenate the word sequences and speaker sequences from all segments. However, the resulting concatenated word sequence may not be identical to the original word sequence from the ASR model due to modifications by LLM. This is undesired and may hurt word error rate. Thus here we need an algorithm to transfer the speaker labels from the concatenated speaker sequence to the original word sequence from the ASR model. We will introduce this algorithm in the following section.

\subsection{Transcript-Preserving Speaker Transfer}
\label{sec:tpst}

Here we describe an algorithm called \emph{Transcript-Preserving Speaker Transfer} (TPST), which will be used in several places in our proposed framework, including training data preparation and the completion parser module.

Assume we have two sets of diarized transcript, referred to as ``source'' and ``target'', each represented by two sequences of the same length: a sequence of words, and a sequence of speaker labels. The purpose of TPST is to transfer the speaker labels from the source sequences to the target sequences, such that:
\begin{enumerate}
    \item The transferred speaker label sequence has a 1-to-1 association with the target word sequence.
    \item The transferred speaker labels are more consistent with the source speaker labels.
\end{enumerate}

As an example, the concatenated word sequence from the completion parser module may not be identical to the original word sequence from the ASR model. Thus we can treat the completion sequences as the source, and the original sequences from the orchestration module as the target, and transfer the speaker labels. Finally, the DiarizationLM outputs will be the original word sequence, associated with the transferred speaker label sequence.

The detailed TPST algorithm is described in Algorithm~\ref{alg:tpst}. An implementation is open sourced on GitHub~\footnote{\url{https://github.com/google/speaker-id/tree/master/DiarizationLM}}.

\begin{algorithm}
\caption{The transcript-preserving speaker transfer (TPST) algorithm.}
\label{alg:tpst}
  \hspace*{\algorithmicindent} \textbf{inputs} \\
  \hspace*{\algorithmicindent}\hspace*{\algorithmicindent} Source word sequence of length $N$: $\mathbf{w}^{src}=(w^{src}_1, \cdots, w^{src}_N)$ \\
  \hspace*{\algorithmicindent}\hspace*{\algorithmicindent} Source speaker sequence of length $N$: $\mathbf{s}^{src}=(s^{src}_1, \cdots, s^{src}_N)$ \\
  \hspace*{\algorithmicindent}\hspace*{\algorithmicindent} Target word sequence of length $M$: $\mathbf{w}^{tgt}=(w^{tgt}_1, \cdots, w^{tgt}_M)$ \\
  \hspace*{\algorithmicindent}\hspace*{\algorithmicindent} Target speaker sequence of length $M$: $\mathbf{s}^{tgt}=(s^{tgt}_1, \cdots, s^{tgt}_M)$ \\
  \\
  \hspace*{\algorithmicindent} \textbf{outputs} \\
  \hspace*{\algorithmicindent}\hspace*{\algorithmicindent} Transferred speaker sequence of length $M$: $\mathbf{s}^{tra}=(s^{tra}_1, \cdots, s^{tra}_M)$ \\
  \begin{algorithmic}[1]
  \Procedure{TPST}{$\mathbf{w}^{src},\mathbf{s}^{src},\mathbf{w}^{tgt},\mathbf{s}^{tgt}$}
  \State Align $\mathbf{w}^{src}$ to $\mathbf{w}^{tgt}$ with the Levenshtein algorithm~\cite{levenshtein1966binary}, resulting in a transform $f_{align}(\cdot)$
  \State $\mathbf{s}^{ali} \gets f_{align}(\mathbf{s}^{src})$
  \qquad \textcolor{gray}{\Comment{$\mathbf{s}^{ali}$ is a speaker sequence of length $M$, and may contain blank speakers $\emptyset$ due to insertion errors in the alignment}}
  \State $K \gets \max \{ \max(\mathbf{s}^{ali}), \max(\mathbf{s}^{tgt}) \}$ \qquad \textcolor{gray}{\Comment{the maximal number of speakers in $\mathbf{s}^{ali}$ and $\mathbf{s}^{tgt}$}}
  \State Initialize a cost matrix $\mathbf{C} \in \mathbb{R}^{K \times K}$
  \For{$1 \leq i \leq K$ and $1 \leq j \leq K$}
    \State $\mathbf{C}_{i,j} \gets \sum_{1 \leq m \leq M} \delta (s_m^{ali}=i \; \textrm{and} \; s_m^{tgt}=j)$
  \EndFor
  \State Solve the assignment problem with cost matrix $\mathbf{C}$ using the Hungarian algorithm~\cite{kuhn1955hungarian}, resulting in a transform $f_{assign}(\cdot)$ \textcolor{gray}{\Comment{handle speaker permutations}}
  \For{$1 \leq m \leq M$}
    \If{$s_m^{ali} \neq \emptyset$} 
        \State $s^{tra}_m \gets f_{assign}(s_m^{ali})$ \textcolor{gray}{\Comment{transfer the speakers from the source}}
    \Else
        \State $s^{tra}_m \gets s_m^{tgt}$ \textcolor{gray}{\Comment{preserve the target speaker if source speaker is unavailable}}
    \EndIf 
  \EndFor
  \EndProcedure
  \end{algorithmic}
\end{algorithm}

Below we show a simple example of the inputs and output of the TPST algorithm:

\begin{tcolorbox}[title=\bf{Example inputs and output of the TPST algorithm}, colback=blue!5!white, colframe=blue!75!black,parbox=false,left=1pt]
\small
\texttt{
\begin{tabular}{l l}
Source words: & hello good morning hi how are you pretty good\\
Source speakers: & 1 1 1 2 2 2 2 1 1\\
Target words: & hello morning hi hey are you be good\\ 
Target speakers: & 1 2 2 2 1 1 2 1\\ 
Transferred speakers: & 1 1 2 2 2 2 1 1
\end{tabular}
}
\end{tcolorbox}

\subsection{LLM finetuning}
\label{sec:finetune}
Although the examples shown in Section~\ref{sec:motivation_example} and Appendix~\ref{sec:other_tasks} were using off-the-shelf Web APIs of commercial LLMs, finetuning the LLM specifically on the speaker diarization task is still required if we need to:
\begin{enumerate}
    \item Reduce errors of a specific speaker diarization system;
    \item Handle more complicated errors;
    \item Keep ASR transcripts unmodified as much as possible from the LLM outputs; 
    \item Avoid undesired leading or tailing text from the generated completions, such as ``Here is the corrected transcript'' or ``We corrected the speakers for these words'';
    \item Use smaller and cheaper LLMs.
\end{enumerate}

To finetune the LLM, we build our training data as a collection of prompt-completion pairs. First, for each utterance, we run the ASR model and the speaker diarization system on it, and apply the orchestration module as shown in Fig.~\ref{fig:diagram}. This will produce the hypothesis word sequence $\mathbf{w}^{hyp}$ and hypothesis speaker sequence $\mathbf{s}^{hyp}$. From the ground truth annotations of this utterance, we build the reference word sequence $\mathbf{w}^{ref}$ and the reference speaker sequence $\mathbf{s}^{ref}$. Given these four sequences, we can build the prompts and completions in our training data with three different flavors, as introduced below.

\subsubsection{Flavor 1: hyp2ora}

The first flavor is named \textit{hypothesis-to-oracle}, or simply \textit{hyp2ora}.
In this flavor, we apply the Transcript-Preserving Speaker Transfer algorithm from Section~\ref{sec:tpst} by treating reference sequences as source and hypothesis sequences as target:
\begin{equation}
    \mathbf{s}^{ora}=\mathrm{TPST}(\mathbf{w}^{ref},\mathbf{s}^{ref},\mathbf{w}^{hyp},\mathbf{s}^{hyp}) ,
\end{equation}
where the output $\mathbf{s}^{ora}$ is the \textbf{oracle hypothesis speakers} transferred from the reference sequences. With $\mathbf{s}^{ora}$, the prompts and completions in our training data are created as below:
\begin{itemize}
    \item \emph{Prompts}: The text representation of $\mathbf{w}^{hyp}$ and $\mathbf{s}^{hyp}$, with segmentation, and optionally prefix and suffix.
    \item \emph{Completions}: The text representation of $\mathbf{w}^{hyp}$ and $\mathbf{s}^{ora}$, with segmentation, and optionally suffix.
\end{itemize}

\subsubsection{Flavor 2: deg2ref}
\label{sec:deg2ref}

The second flavor is named \textit{degraded-to-reference}, or simply \textit{deg2ref}.
In this flavor, we apply the Transcript-Preserving Speaker Transfer algorithm from Section~\ref{sec:tpst} by treating hypothesis sequences as source and reference sequences as target:
\begin{equation}
    \mathbf{s}^{deg}=\mathrm{TPST}(\mathbf{w}^{hyp},\mathbf{s}^{hyp},\mathbf{w}^{ref},\mathbf{s}^{ref}) ,
\end{equation}
where the output $\mathbf{s}^{deg}$ is the \textbf{degraded reference speakers} transferred from the hypothesis sequences. With $\mathbf{s}^{deg}$, the prompts and completions in our training data are created as below:
\begin{itemize}
    \item \emph{Prompts}: The text representation of $\mathbf{w}^{ref}$ and $\mathbf{s}^{deg}$, with segmentation, and optionally prefix and suffix.
    \item \emph{Completions}: The text representation of $\mathbf{w}^{ref}$ and $\mathbf{s}^{ref}$, with segmentation, and optionally suffix.
\end{itemize}

\subsubsection{Flavor 3: mixed}

The third flavor named \textit{mixed} is simply the union of the prompts and completions from the previous two flavors. When building training batches, prompt-completion pairs from the two flavors are interleaved.

Note that for all three flavors, it is critical for the prompt and completion to use the same word sequence with different speaker sequences. This helps the LLM to focus on correcting the speaker labels without modifying the ASR transcripts.

\section{Experiments}
\label{sec:exp}

\subsection{Datasets}
\label{sec:data}
To finetune the LLM, we use the training subset of the Fisher corpus~\cite{cieri2004fisher}, which consists of 1,920 hours of 11,527 conversations. The same train-test split of the Fisher dataset has been used in many previous works~\cite{zhao2022augmenting,wang2022highly,paturi2023lexical,zhao2023usm}

For evaluation, we use the testing subset of the Fisher corpus~\cite{cieri2004fisher}, as well as the testing subset of
Callhome American English data~\cite{canavan1997callhome}. The Fisher testing subset consists of 28.7 hours of 172 conversations\footnote{\url{https://github.com/google/speaker-id/blob/master/publications/ScdLoss/eval/fisher.txt}}. The Callhome American English testing subset consists of 1.7 hours of 20 conversations. Both datasets are in the telephone speech domain, and all conversations have 2 speakers.

\subsection{Metrics}
To evaluate the diarization performance, we use two metrics: the
Word Diarization Error Rate (WDER)~\cite{shafey2019joint}
and the 
concatenated minimum-permutation word error rate (cpWER)~\cite{watanabe2020chime}.
To briefly recap, WDER is defined as:
\begin{equation}
    \mathrm{WDER}=\frac{S_\mathrm{IS}+C_\mathrm{IS}}{S+C},
\end{equation}
where,
\begin{enumerate}
    \item $S_\mathrm{IS}$ is the number of ASR Substitutions with Incorrect Speaker tokens.
    \item $C_\mathrm{IS}$ is the number of Correct ASR words with Incorrect Speaker tokens.
    \item $S$ is the number of ASR substitutions.
    \item $C$ is the number of Correct ASR words.
\end{enumerate}

And cpWER is computed as follows:
\begin{enumerate}
    \item Concatenate all transcripts of each speaker for both reference and hypothesis.
    \item Compute the WER between the reference and all possible speaker permutations of the hypothesis. 
    \item Pick the lowest WER among all these permutations, which is assumed to be the best permutation.
\end{enumerate}

All three metrics reported in this paper (WER, WDER, and cpWER) are micro metrics, i.e. both numerators and denominators are aggregated on the entire dataset.

\subsection{Models}
\begin{figure}
	\centering
	\includegraphics[width=0.7\linewidth]{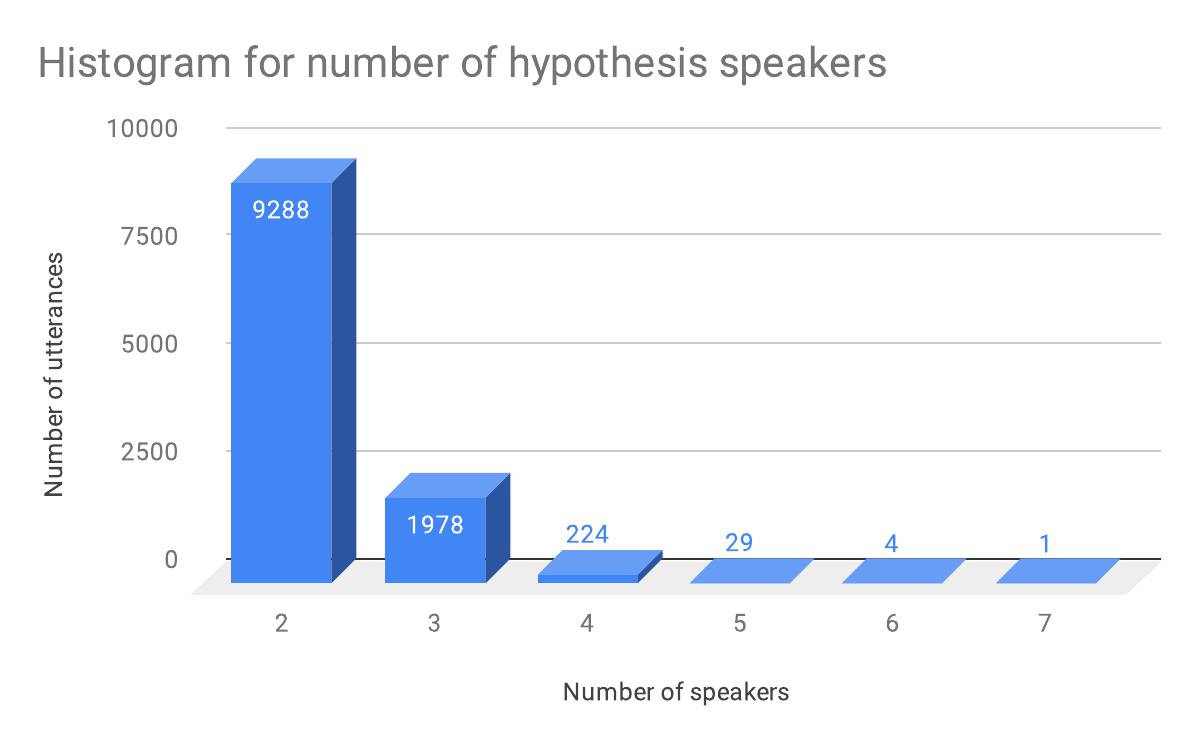}
	\caption{The histogram for the number of hypothesis speakers predicted by the turn-to-diarize system on the Fisher training set. Note that the ground truth number of speakers is always two on the Fisher dataset, but we do not constrain the number of speakers for the turn-to-diarize system.}
	\label{fig:speaker_histogram}
\end{figure}

For the ASR model in Fig.~\ref{fig:diagram}, we use a universal speech model (USM) ~\cite{zhang2023google} with 600 million parameters trained with the RNN-T loss~\cite{graves2012sequence}. For the speaker diarization model in Fig.~\ref{fig:diagram}, we use the turn-to-diarize system~\cite{xia2022turn} with a multi-stage clustering setup~\cite{wang2022highly} in our experiments, which is capable of diarizing hours of audio recordings in real time on a mobile device~\cite{speakerlabelsblog}. The number of speakers is unknown (from $1$ to $\infty$) to the speaker diarization system in all of our experiments.
Specifically, for the Fisher training set, the Fisher testing set, and the Callhome testing set, all utterances have a ground truth number of speakers equal to two, but our turn-to-diarize system may predict any number of hypothesis speakers. The histogram of number of hypothesis speakers predicted by turn-to-diarize on the Fisher training set is shown in Figure~\ref{fig:speaker_histogram}.  
Although our experiments are based on the USM + turn-to-diarize setup, we would like to point out that the proposed framework is very generic and should work with other ASR or speaker diarization systems as well, such as variants of end-to-end speaker diarization models~\cite{fujita2019end,ueda2022eend,e2ediarizationpatent,horiguchi2020end}.

For the LLM in Fig.~\ref{fig:diagram}, we experiment with the PaLM 2-S model (``\texttt{text-bison}'' model in Google Cloud API) and the PaLM 2-L model (``\texttt{text-unicorn}'' model in Google Cloud API)~\cite{palm2}.
We use the PaLM 2-S model as our foundation model, and finetune it on the dataset described in Section~\ref{sec:data} with data processing steps described in Section~\ref{sec:finetune}. This model uses a sentence piece model (SPM) of 256k tokens as its tokenizer~\cite{kudo2018sentencepiece}.
During finetuning, we limit the LLM input size by 4,096 tokens, and segment our training and testing data accordingly. The PaLM 2-L model will only be used for zero-shot and one-shot experiments, as described in Section~\ref{sec:0shot-1shot}. We also experimented with open source models such as Llama 2~\cite{touvron2023llama} and Llama 3. Model details on how we finetune these models are provided in Appendix~\ref{sec:open_source_models}.

In our prompt builder module, we use an empty prompt prefix, and a 5-character prompt suffix ``\texttt{\textvisiblespace-{}->\textvisiblespace}'' (note the two spaces around the arrow). For the completions in our training data, we use a 6-character completion suffix ``\texttt{\textvisiblespace[eod]}'' (short for ``end of document''; note the leading space).
After processing the training data with the prompt builder module, we result in 13,430 prompt-completion pairs for training in total. The average length of a prompt is 2,371 SPM tokens, and the average length of a completion is 2,329 tokens. The LLM is trained for 1,200 steps with a batch size of 16.

\subsection{Zero-shot and one-shot baselines}
\label{sec:0shot-1shot}
Apart from finetuning the PaLM 2-S model on the speaker diarization task, we also experiment with directly using the PaLM 2-S and PaLM 2-L models on the speaker diarization task without finetuning. This is more similar to the example we demonstrated in Section~\ref{sec:motivation_example}.

For the zero-shot setup, we use a prompt prefix that contains an instruction describing the task, as shown below.

\begin{tcolorbox}[title=\bf{Prompt prefix for zero-shot}, colback=blue!5!white, colframe=blue!75!black,parbox=false]
In the speaker diarization transcript below, some words are potentially misplaced.

Please correct those words and move them to the right speaker.

Directly show the corrected transcript without explaining what changes were made or why you made those changes.\textbackslash n
\end{tcolorbox}

For the one-shot setup, the prompt prefix contains both the instruction describing the task, and also a small example, as shown below.

\begin{tcolorbox}[title=\bf{Prompt prefix for one-shot}, colback=blue!5!white, colframe=blue!75!black,parbox=false]
In the speaker diarization transcript below, some words are potentially misplaced.
Please correct those words and move them to the right speaker. For example, given this input transcript,

<spk:1> How are you doing today? I
<spk:2> am doing very well. How was everything at the
<spk:1> party? Oh, the party? It was awesome. We had lots of fun. Good
<spk:2> to hear!

The correct output transcript should be:

<spk:1> How are you doing today?
<spk:2> I am doing very well. How was everything at the party?
<spk:1> Oh, the party? It was awesome. We had lots of fun. <spk:2> Good to hear!

Now, please correct the transcript below.\textbackslash n
\end{tcolorbox}

\subsection{Evaluation results}

In Table~\ref{table:results}, we show the evaluation results of the USM + turn-to-diarize baseline together with the outputs post-processed by DiarizationLM. We report results for zero-shot, one-shot, and finetuning on the diarization task with three different flavors.

For zero-shot and one-shot experiments with PaLM 2-S, we observe significantly worse WDER and cpWER performance compared with the baseline system, indicating the PaLM 2-S foundation model does not offer speaker diarization capabilities without finetuning. Zero-shot experiment with PaLM 2-L model also shows bad performance, while one-shot experiment with PaLM 2-L model is much better, but still worse than the baseline system. Our results indicate that the PaLM 2-L model with one-shot is able to improve speaker diarization in relatively simple cases as shown in Section~\ref{sec:motivation_example} and Appendix~\ref{sec:other_tasks}.
However, real world applications can be much more complicated with errors from both the ASR system and the speaker diarization system. In such cases, even with one-shot, LLM can still introduce even more errors to the results if not finetuned specifically on the speaker diarization task. 

On both datasets, we observe big improvement of both WDER and cpWER with any of the three finetuning flavors.
Interesting, the biggest improvement is observed with the hyp2ora flavor, while the smallest improvement is observed with the deg2ref flavor.
Specifically for hyp2ora, we see a rel. 55.5\% improvement of WDER after post-processing with DiarizationLM on the Fisher testing set. Even if we did not use any Callhome data during the LLM finetuning, we see a rel. 44.9\% improvement of WDER on the Callhome testing set. The WER of the USM on the two testing sets are relatively high due to domain mismatch and suboptimal annotation quality of the ground truth. However, this also demonstrated that the DiarizationLM solution provides consistent quality gains even with out-of-domain ASR and speaker diarization models.

To further demonstrate this, in Table~\ref{table:results_with_ref}, we show the results of a similar setup, but we replace the USM-based ASR model directly by the ground truth ASR transcripts from the testing sets. For these experiments, we will have WER=0\%, and the hyp2ora and deg2ref flavors will be equivalent. From the table, we can still see big improvements of WDER after post-processing the diarization results by the same DiarizationLM model (i.e. deg2ref flavor in Table~\ref{table:results}).

\subsection{Case studies}

Based on the results from Table~\ref{table:results}, we also present example cases from the Fisher and Callhome testing sets where we see big improvements of WDER in Table~\ref{table:fisher_examples} and Table~\ref{table:callhome_examples}, respectively. From these examples, we are seeing multiple patterns of corrections:
\begin{itemize}
    \item DiarizationLM make corrections where \textbf{different parts of sentence} are moved to the same speaker, e.g. ``it's more of'' and ``it'll be warm'' in \texttt{fe\_03\_07146} from Table~\ref{table:fisher_examples}. This is consistent with our initial observations as demonstrated in Section~\ref{sec:motivation_example}.
    \item DiarizationLM can merge short speaker turns due to \textbf{disfluency}, such as ``yeah yeah'' and  ``i i hear i hear '' in \texttt{fe\_03\_11159} from Table.~\ref{table:fisher_examples}. Diarization errors from disfluency usually attribute to low quality speaker embeddings extracted from very short speaker turn segments.
    \item DiarizationLM can also detect speaker turns due to \textbf{interruptions}, such as ``oh all right'' in \texttt{fe\_03\_11210} from Table~\ref{table:fisher_examples}, and ``oh my'' in \texttt{en\_6408} from Table~\ref{table:callhome_examples}.
\end{itemize}

\begin{table}
    \centering
    \caption{Evaluation results of the USM + turn-to-diarize baseline system and the results post-processed by DiarizationLM. For DiarizationLM, we experiment with PaLM 2 foundation models with and without finetuning on the diarization task. We also experiment with Llama 2 and Llama 3 models finetuned on the diarization task (model details in Appendix~\ref{sec:open_source_models}). WERs are the same for all systems due to TPST. All numbers are percentages.}
    \begin{tabular}{l|c r r|c r r}
    \toprule
    \multirow{2}{*}{System} & \multicolumn{3}{c|}{Fisher testing set} & \multicolumn{3}{c}{Callhome testing set}   \\
    & WER & WDER & cpWER & WER & WDER & cpWER \\ \hline
    USM + turn-to-diarize baseline & 15.48 & 5.32 &21.19 & 15.36 & 7.72 & 24.39 \\
    + PaLM 2-S zero-shot & - & 11.96 & 30.19 & - & 12.26 & 30.60 \\
    + PaLM 2-S one-shot & - & 16.58 & 38.03 & - & 14.50 & 34.32 \\
    + PaLM 2-L zero-shot & - & 11.36 & 31.78 & - & 13.29 & 34.30 \\
    + PaLM 2-L one-shot & - & 5.94 & 22.21 & - & 7.95 & 24.67 \\
    + PaLM 2-S finetuned (hyp2ora)  & - & \boldgreen{2.37} & \boldgreen{16.93} & - & \boldgreen{4.25} & \boldgreen{20.22} \\
    + PaLM 2-S finetuned (deg2ref)  & - & 3.94 & 18.55 & - & 5.33 & 21.47 \\
    + PaLM 2-S finetuned (mixed)  & - & 2.41 & 16.94 & - & 4.76 & 20.84 \\ \hline
    + Llama 2 13B finetuned (mixed) v1 & - & 3.65 & 18.92 & - & 8.02 & 25.11  \\
    + Llama 3 8B finetuned (mixed) v1 & - & 4.40 & 19.76 & - & 12.27 & 30.80  \\
    + Llama 3 8B finetuned (mixed) v2 & - & 3.28 & 18.37 & - & 6.66 & 23.57  \\
    \bottomrule 
    \end{tabular}
    \label{table:results}
\end{table}

\begin{table}
    \centering
    \caption{Evaluation results of the turn-to-diarize baseline system with reference ASR transcript (assuming WER=0\%) and the results post-processed by DiarizationLM. For DiarizationLM, we experiment with PaLM 2 foundation models with and without finetuning on the diarization task. All numbers are percentages.}
    \begin{tabular}{l|r r|r r}
    \toprule
    \multirow{2}{*}{System} & \multicolumn{2}{c|}{Fisher testing set} & \multicolumn{2}{c}{Callhome testing set}   \\
    &  WDER & cpWER & WDER & cpWER \\ \hline
    Reference + turn-to-diarize baseline  & 2.81 & 5.19  & 3.74 & 6.82 \\
    + PaLM 2-S zero-shot & 7.50 & 12.70 & 7.29 & 12.79 \\
    + PaLM 2-S one-shot & 10.92 & 19.16 & 12.79 & 21.65 \\
    + PaLM 2-L zero-shot & 8.69 & 16.85  & 11.67 & 22.87 \\
    + PaLM 2-L one-shot & 3.23 & 5.99 & 3.76 & 6.95 \\
    + PaLM 2-S finetuned  & \boldgreen{1.18} & \boldgreen{2.21} & \boldgreen{1.49} & \boldgreen{2.66} \\
    \bottomrule 
    \end{tabular}
    \label{table:results_with_ref}
\end{table}

We also look into why zero-shot and one-shot experiments in Table~\ref{table:results} produced worse results than the baseline system. We found that without finetuning on the speaker diarization tasks, zero-shot and one-shot outputs from the LLM often delete big chunks of hypothesis text from the prompt. Finetuning the LLM is critical to avoid such undesired deletions. A few zero-shot examples with the PaLM 2-S model from the Fisher testing set were shown in Table~\ref{table:zero_shot_examples}.

\begin{table}
\small
    \centering
    \caption{Example cases from the Fisher testing set where we see big absolute WDER reduction ($\Delta$ WDER) with DiarizationLM (deg2ref flavor).}
    \begin{tabularx}{\textwidth}{>{\hsize=.12\hsize}X|>{\hsize=.44\hsize}X|>{\hsize=.44\hsize}X} 
     \toprule
     \textbf{Utterance} & \textbf{Before DiarizationLM} & \textbf{After DiarizationLM} \\
     \hline
     fe\_03\_07146 \newline ($\Delta$ WDER \newline =8.80\%) &
...\newline<spk:3> it's it's 
\newline<spk:1> more of summer always like you know we never experience a bit cold over here 
\newline<spk:4> usually it'll 
\newline<spk:1> be warm or like very hot in summer yeah and 
\newline<spk:3> extremely hot yeah with high humidity my humidity is pretty 
\newline<spk:1> much high because i stay close to the sea coast over here 
\newline<spk:3> yeah 
\newline<spk:1> so 
\newline<spk:3> that makes you live houston is it like houston where you live yeah i i i live 
\newline<spk:1> in houston
\newline...
& 
...\newline<spk:1> it's it's more of summer always like you know we never experience a bit cold over here usually it'll be warm or like very hot in summer 
\newline<spk:2> yeah and extremely hot yeah with high humidity my 
\newline<spk:1> humidity is pretty much high because i stay close to the sea coast over here 
\newline<spk:2> yeah so that makes you live houston is it like houston where you live 
\newline<spk:1> yeah i i i live in houston
\newline...\\ 
     \hline
     fe\_03\_06816 \newline ($\Delta$ WDER \newline =6.61\%) & 
...\newline<spk:3> uhuh 
\newline<spk:2> did you see the the woman golfer that was on this the one 
\newline<spk:1> monica yeah yeah 
\newline<spk:2> what's her name monica stone yeah mhm she she 
\newline<spk:1> blew out she fell out of that tournament but i didn't think she'd do it she she's girls can't compete against guys  \newline...
&
...\newline<spk:2> uhuh did you see the the woman golfer that was on this the one 
\newline<spk:1> monica yeah yeah 
\newline<spk:2> what's her name monica stone 
\newline<spk:1> yeah 
\newline<spk:2> mhm 
\newline<spk:1> she she blew out she fell out of that tournament but i didn't think she'd do it she she's girls can't compete against guys 
\newline...\\
     \hline
     fe\_03\_11210 \newline ($\Delta$ WDER  \newline =6.35\%) & 
...
\newline<spk:1> the vikings mine's the eagles i'm from new jersey oh all right i have my jersey on now i watch the game tonight yeah well i i may i may just watch
\newline<spk:2> part of it tonight too then but uh it's a case as i say if if i had to pay for it i probably wouldn't watch it
\newline<spk:1> i wouldn't either uhhuh
\newline<spk:2> unless
\newline<spk:1> it was an eagles game
\newline... 
&
...
\newline<spk:1> the vikings mine's the eagles i'm from new jersey
\newline<spk:2> oh all right
\newline<spk:1> i have my jersey on now i watch the game tonight yeah
\newline<spk:2> well i i may i may just watch part of it tonight too then but uh it's a case as i say if if i had to pay for it i probably wouldn't watch it
\newline<spk:1> i wouldn't either
\newline<spk:2> uhhuh
\newline<spk:1> unless it was an eagles game
\newline...\\
   \hline
     fe\_03\_11159 \newline ($\Delta$ WDER  \newline =4.05\%) & 
...
\newline<spk:2> yeah 
\newline<spk:1> anniversary that's horrible 
\newline<spk:2> yeah 
\newline<spk:1> yeah it's not good 
\newline<spk:2> i 
\newline<spk:1> i hear i hear you there that's not a good thing you 
\newline<spk:2> know i mean of course you know that's a day that will go down instantly nobody will ever remember it 
\newline...
&
...
\newline<spk:1> yeah anniversary that's horrible yeah yeah it's not good i i hear i hear you there that's not a good thing 
\newline<spk:2> you know i mean of course you know that's a day that will go down instantly nobody will ever remember it
\newline...\\
     \bottomrule
    \end{tabularx}
    \label{table:fisher_examples}
\end{table}

\begin{table}
\small
    \centering
    \caption{Example cases from the Callhome testing set where we see big absolute WDER reduction ($\Delta$ WDER) with DiarizationLM (deg2ref flavor).}
    \begin{tabularx}{\textwidth}{>{\hsize=.12\hsize}X|>{\hsize=.44\hsize}X|>{\hsize=.44\hsize}X} 
     \toprule
     \textbf{Utterance} & \textbf{Before DiarizationLM} & \textbf{After DiarizationLM} \\
     \hline
     en\_6447 \newline($\Delta$ WDER \newline=12.49\%) & 
...\newline<spk:1> i'm 
\newline<spk:2> going to see if i can talk to the guy that's selling the trailer if i can chew him down a bit uhhuh 
\newline<spk:1> and 
\newline<spk:2> you know what you just said benedicta is are you living with benedicta 
\newline<spk:1> yes yes yes 
\newline<spk:2> you know what i bet she answered the phone 
\newline... 
& 
...\newline<spk:2> i'm going to see if i can talk to the guy that's selling the trailer if i can chew him down a bit 
\newline<spk:1> uhhuh 
\newline<spk:2> and you know what you just said benedicta is are you living with benedicta 
\newline<spk:1> yes yes yes 
\newline<spk:2> you know what i bet she answered the phone 
\newline... \\ 
     \hline
     en\_6408 \newline ($\Delta$ WDER \newline =10.87\%) &  
...\newline<spk:1> uhu 
\newline<spk:2> so 
\newline<spk:1> he had big surgery again and he's in a wheelchair oh my 
\newline<spk:2> and 
\newline<spk:1> he doesn't want to go to school in a wheelchair uhuh but 
\newline<spk:2> he might he wants to have tutoring at home but they're still where they lived on 45th street 
\newline<spk:1> yeah they're there \newline... 
& 
...\newline<spk:2> uhu 
\newline<spk:1> so he had big surgery again and he's in a wheelchair 
\newline<spk:2> oh my 
\newline<spk:1> and he doesn't want to go to school in a wheelchair 
\newline<spk:2> uhuh 
\newline<spk:1> but he might he wants to have tutoring at home 
\newline<spk:2> but they're still where they lived on 45th street 
\newline<spk:1> yeah they're there \newline...\\ 
     \hline
     en\_6298 \newline ($\Delta$ WDER \newline =9.95\%) &  
...\newline<spk:1> um hey we're we're confused about you guys address
\newline<spk:2> is
\newline<spk:1> it 1324 or 13
\newline<spk:2> it's 1 324
\newline<spk:1> excuse me 1324 yes and it's me view is me two words or one word yes it's two words and there's an ln besides\newline... 
& 
...\newline<spk:1> um hey we're we're confused about you guys address is it 1324 or 13
\newline<spk:2> it's 1 324
\newline<spk:1> excuse me 1324
\newline<spk:2> yes
\newline<spk:1> and it's me view is me two words or one word
\newline<spk:2> yes it's two words and there's an ln besides\newline...\\ 
\hline
     en\_4792 \newline ($\Delta$ WDER \newline =9.42\%) &  
...
\newline<spk:2> yeah well he was at columbia 
\newline<spk:1> he was there like five years and they turned him down for tenure then he went somewhere else he he was down in college park maryland yeah and he i think he was only non tenure track down there then supposedly supposed to be back in japan now yeah but you know he's he's probably become an english teacher at some unit yeah i know a guy believe it or not i know a guy from manhattan who was up in sapotto his major he did an mba believe it or not he's he's an english teacher now huh 
\newline...
&
...
\newline<spk:2> yeah well he was at columbia 
\newline<spk:1> he was there like five years and they turned him down for tenure then he went somewhere else he he was down in college park maryland 
\newline<spk:2> yeah 
\newline<spk:1> and he i think he was only non tenure track down there then supposedly supposed to be back in japan now 
\newline<spk:2> yeah 
\newline<spk:1> but you know he's he's probably become an english teacher at some unit 
\newline<spk:2> yeah 
\newline<spk:1> i know a guy believe it or not i know a guy from manhattan who was up in sapotto his major he did an mba believe it or not he's he's an english teacher now 
\newline<spk:2> huh 
\newline...
     \\
     \bottomrule
    \end{tabularx}
    \label{table:callhome_examples}
\end{table}

\begin{table}
\small
    \centering
    \caption{Example cases from the Fisher testing set where zero-shot PaLM 2-S deletes lots of text from the prompt.}
    \begin{tabularx}{\textwidth}{>{\hsize=.12\hsize}X|>{\hsize=.44\hsize}X|>{\hsize=.44\hsize}X} 
     \toprule
     \textbf{Utterance} & \textbf{Before DiarizationLM} & \textbf{After DiarizationLM} \\
     \hline
     fe\_03\_11252 & ...
\newline<spk:1> oh okay i believe it's a lot wrong with the public schools i don't believe that they're um that they're giving um these kids a sense of um well they're not teaching them what they need to know once they get out of um school you know mhm um what what's happening is that's probably why you got a lot of um a lot of people that's unemployed i think you know they you get a lot from school and they taking a lot of um i guess the economics out of school you know
\newline<spk:2> right
\newline...
 & ...
\newline<spk:1> oh okay i believe it's a lot wrong with the public schools i don't believe that they're um that they're giving um these kids a sense of um well they're not teaching them what they need to know once they get out of um school you know
\newline<spk:2> right
\newline...\\ \hline
fe\_03\_11224
& ...
\newline<spk:1> so um i think what do you think is an important thing in a relation i think the topic was um what you um what are the most important things in a life partner yeah uh h well what do you think me 
\newline<spk:2> i would have to say trust and honesty like cuz without that you really don't have nothing to build on you know right yeah 
\newline...
& ...
\newline<spk:1> so um i think what do you think is an important thing in a relation
\newline<spk:2> i would have to say trust and honesty like cuz without that you really don't have nothing to build on you know right
     \newline...\\
     \bottomrule
    \end{tabularx}
    \label{table:zero_shot_examples}
\end{table}

\section{Discussion and future work}

The experiments in Section~\ref{sec:exp} have shown very promising results where LLMs can significantly reduce speaker diarization errors. However, we also admit the limitations of these experiments. First of all, the training and testing data from the experiments are all based on the telephone speech domain, all with exactly 2 speakers.
An important future work would be to include more diverse datasets to finetune the LLM, and evaluate its performance across different domains with unknown number of speakers.

In Appendix~\ref{sec:other_tasks}, we have demonstrated other diarization capabilities of LLMs. However, due to lack of relevant datasets, we haven't been able to thoroughly evaluate these capabilities. One interesting future work would be to collect datasets of these tasks and evaluate how LLM performs.

Another research direction would be to compare different LLMs, in different size variants on the speaker diarization task. Specifically, the performance will likely be even better if we finetune larger models such as PaLM 2-M or PaLM 2-L. It would also be interesting to reproduce the experiments with other speaker diarization systems such as EEND~\cite{fujita2019end} or WEEND~\cite{huang24d_interspeech}.

Lastly, as PaLM 2 models are multilingual~\cite{palm2}, the DiarizationLM framework can naturally apply to speaker diarization tasks in other languages. It would be helpful to evaluate how DiarizationLM performs on speaker diarization datasets in other languages than English.

\section{Related work}

\subsection{Speaker diarization post-processing}

In the context of conventional speaker diarization, ``post-processing'' usually refers to a stage where the clustering results are refined with signals from other sources or systems. An early post-processing approach was known as ``resegmentation'', where the Gaussian mixture models (GMMs) are estimated for each speaker with the Baum-Welch algorithm, and a Viterbi algorithm is used to re-annotate the speakers with the GMMs~\cite{sell2015diarization}. Later in ~\cite{yin2018neural}, the authors proposed to use a neural network for resegmentation, with an additional class for non-speech.
In~\cite{han2023diacorrect}, the authors proposed DiaCorrect, a method inspired by error correction techniques in ASR. DiaCorrect uses parallel convolutional encoders for the speakers from the initial diarization results and a transformer based decoder to produce corrected diarization results.
One major difference in our proposed framework is that we leverage semantic information to refine the diarization results on a word level, while these resegmentation approaches are only based on acoustic information and perform at cluster level.

Another type of post-processing is to combine the outputs of multiple speaker diarization systems, e.g. via majority voting~\cite{huijbregts2009majority}, speaker matching~\cite{bozonnet2010system}, or both~\cite{stolcke2019dover}. More recently in~\cite{park2021multi}, the authors proposed to perform speaker diarization on different temporal scales, and combine their outputs via 1-D convolutional neural networks. In~\cite{horiguchi2021end}, the authors proposed to use end-to-end speaker diarization as a post-processing step for initial speaker diarization results of a clustering-based system. Our proposed framework is generic such that it can apply to either the results of a single speaker diarization system, or to the combined results of multiple speaker diarization systems.

\subsection{Speaker diarization with semantic information}
Apart from the joint ASR and speaker diarization models discussed in Section~\ref{sec:intro},
researchers have also studied various approaches of integrating semantic information into conventional speaker diarization systems. Some of the benefits of DiarizationLM may also be achieved with non-LLM methods.

The most common approach to leverage semantic information is to use ASR word alignments to refine the voice activity detection and initial segmentation~\cite{silovsky2012incorporation}. A variant of this approach is to build a speaker turn detection model and segment by speaker turns~\cite{park2018multimodal}. In~\cite{park2020speaker}, a Gated Recurrent Units (GRUs)~\cite{chung2014empirical} based speaker turn probability estimator is trained on top of word embeddings and speaker embeddings, and the estimated probabilities are combined with the adjacency matrix for spectral clustering. Similarly in~\cite{xia2022turn}, an end-to-end trained transformer transducer (T-T)~\cite{zhang2020transformer} based speaker turn detection model is used to constrain the spectral clustering via Exhaustive and Efficient Constraint Propagation (E2CP).

\subsection{Speaker diarization with LLM}

In ~\cite{paturi2023lexical}, the authors proposed Speaker Error Corrector (SEC), which aims to solve the same problem as we stated in Section~\ref{sec:intro}. In ~\cite{paturi2023lexical}, word embeddings from the ASR transcript are extracted with a pre-trained Roberta-base LM~\cite{liu2019roberta}. Then a separately trained transformer encoder takes the word embeddings and the hypothesis speaker labels as input, and produces the corrected speaker labels. The transformer encoder is trained on both simulated diarization errors and real data. The biggest difference from our proposed framework to~\cite{paturi2023lexical} is that we directly feed the compact pure textual representation of the ASR and diarization results as part of the prompt to the LLM, and directly finetune the LLM to produce the corrected results in the same compact textual representation. Our DiarizationLM is a ``text-in, text-out'' system, without relying on internal embedding representations from the LLM.

More recently in ~\cite{park2023enhancing}, the authors proposed to use LLM to predict the speaker probability for the next word, and incorporate this probability into the beam search decoding of speaker diarization. Our proposed framework differs from this work by using a single prompt (or several prompts due to LLM input size limit) to post-process the entire results of the speaker diarization system, instead of word-by-word prompting. Additionally, our proposed framework can be more generally applied to any speaker diarization system, instead of requiring word-level speaker probabilities for beam search decoding.

\section{Conclusion}

In this paper, we demonstrate that large language models (LLM) can be used to post-process speaker diarization results, achieving various goals such as improving the readability of the diarization transcript, and reducing the diarization errors. Specifically, we proposed DiarizationLM, a framework where we use a finetuned LLM to refine the results from off-the-shelf ASR and speaker diarization systems. We introduced three different flavors to build the prompt-completion pairs data for finetuning the LLM. Our experiments on Fisher and Callhome datasets show that a finetuned PaLM 2-S model can drastically reduce the word diarization error rates of typical diarization systems like turn-to-diarize.

\bibliographystyle{IEEEbib}
\bibliography{refs}

\begin{appendices}

\section{Open source models}
\label{sec:open_source_models}

Apart from PaLM 2~\cite{palm2} models, we have also experimented with open source models such as Llama 2~\cite{touvron2023llama} and Llama 3. These finetuned models are publicly available on Hugging Face. The finetuning scripts are available at: \url{https://github.com/google/speaker-id/tree/master/DiarizationLM/unsloth}.

\subsection{\texttt{\textup{google/DiarizationLM-13b-Fisher-v1}}}
\label{sec:llama2_13b_fisher_v1}

\begin{tabularx}{\textwidth}{l X} 
Model URL: & \url{https://huggingface.co/google/DiarizationLM-13b-Fisher-v1}  \\ 
Foundation model: & \url{https://huggingface.co/unsloth/llama-2-13b-bnb-4bit}
\end{tabularx}

This model is finetuned on the training subset of the Fisher corpus, using a LoRA adapter~\cite{hu2021lora} of rank 256. The total number of training parameters is 1,001,390,080. With a batch size of 16, this model has been trained for 12000 steps, which is $\sim$4 epochs of the training data.

We use the ``mixed'' flavor during our training, meaning we combine data from ``hyp2ora'' and ``deg2ref'' flavors (see Section~\ref{sec:finetune} for the definition of these flavors). After the prompt builder, we have a total of 48,142 prompt-completion pairs in our training set.

The finetuning took more than 3 days on a Google Cloud VM instance that has one NVIDIA A100 GPU with 80GB memory.

The maximal length of the prompt to this model is 6000 characters, including the ``\texttt{\textvisiblespace-{}->\textvisiblespace}'' suffix. The maximal sequence length is 4096 tokens.

The evaluation results of this model is also provided in Table~\ref{table:results}. We can see that similar to the finetuned PaLM 2-S models, this model also significantly reduces the WDER and cpWER on the Fisher testing set. However, on the Callhome testing set, we are seeing degradation in WDER and cpWER, indicating overfitting on the Fisher dataset. We have found this is because the loss is computed on both the prompt and the completion tokens during training.

\subsection{\texttt{\textup{google/DiarizationLM-8b-Fisher-v1}}}
\label{sec:llama3_8b_fisher_v1}

\begin{tabularx}{\textwidth}{l X} 
Model URL: & \url{https://huggingface.co/google/DiarizationLM-8b-Fisher-v1}  \\ 
Foundation model: & \url{https://huggingface.co/unsloth/llama-3-8b-bnb-4bit}
\end{tabularx}

This model is finetuned on the training subset of the Fisher corpus, using a LoRA adapter~\cite{hu2021lora} of rank 256. The total number of training parameters is 671,088,640. With a batch size of 16, this model has been trained for 25400 steps, which is $\sim$8 epochs of the training data.

Other configurations of this model is the same as the model in the previous section. Note that we still use a maximal sequence length of 4096 tokens even if Llama 3 supports longer sequence.

The finetuning took more than 4 days on a Google Cloud VM instance that has one NVIDIA A100 GPU with 80GB memory.

The evaluation results of this model is also provided in Table~\ref{table:results}. We can see that its performance is significantly worse than the Llama 2 13B model from Appendix~\ref{sec:llama2_13b_fisher_v1}, but still better than the USM + turn-to-diarize baseline on Fisher testing set. Similarly, the performance on Callhome is worse than baseline, because the loss is computed on both the prompt and the completion tokens during training.

\subsection{\texttt{\textup{google/DiarizationLM-8b-Fisher-v2}}}
\label{sec:llama3_8b_fisher_v2}

\begin{tabularx}{\textwidth}{l X} 
Model URL: & \url{https://huggingface.co/google/DiarizationLM-8b-Fisher-v1}  \\ 
Foundation model: & \url{https://huggingface.co/unsloth/llama-3-8b-bnb-4bit}
\end{tabularx}

This model is finetuned on the training subset of the Fisher corpus, using a LoRA adapter~\cite{hu2021lora} of rank 256. The total number of training parameters is 671,088,640. With a batch size of 16, this model has been trained for 28800 steps, which is $\sim$9 epochs of the training data.

Note that when training this model, \textbf{the loss is only computed on the completion tokens}. This is the biggest different between this model and the previous model in Appendix~\ref{sec:llama3_8b_fisher_v1} ( \texttt{google/DiarizationLM-8b-Fisher-v1}).

Other configurations of this model is the same as the model in the previous section. Note that we still use a maximal sequence length of 4096 tokens even if Llama 3 supports longer sequence.

The finetuning took more than 4 days on a Google Cloud VM instance that has one NVIDIA A100 GPU with 80GB memory.

The evaluation results of this model is also provided in Table~\ref{table:results}. We can see that its performance is significantly better than the model from Appendix~\ref{sec:llama3_8b_fisher_v1}. Specifically, we no longer see degradation on Callhome testin set.

\subsection{\texttt{\textup{diarizers-community/DiarizationLM-Gemma-4-E4B-v1}}}
\label{sec:gemma4_e4b_v1}

\begin{tabularx}{\textwidth}{l X} 
Model URL: & \url{https://huggingface.co/diarizers-community/DiarizationLM-Gemma-4-E4B-v1}  \\ 
Foundation model: & \url{https://huggingface.co/google/gemma-4-E4B}
\end{tabularx}

This model is finetuned on a multi-domain mixture combining the training subsets of four speaker diarization corpora---Fisher~\cite{cieri2004fisher}, Callhome, ICSI~\cite{janin2003icsi}, and AMI~\cite{carletta2005ami}---built on top of the 4B-parameter dense Gemma 4 E4B foundation model~\cite{gemma4_2026}. We apply a LoRA adapter~\cite{hu2021lora} of rank 256 across all attention projections (\texttt{q\_einsum}, \texttt{kv\_einsum}, \texttt{attn\_vec\_einsum}), feed-forward layers (\texttt{gating\_einsum}, \texttt{linear}), and per-layer input projections. Consistent with Appendix~\ref{sec:llama3_8b_fisher_v2}, \textbf{the cross-entropy training loss is computed strictly on the completion tokens}.

To enable simultaneous improvements on both 2-speaker telephone conversations (Fisher, Callhome) and multi-speaker tabletop meetings (ICSI, AMI) without requiring any modifications to standard inference-time Transcript-Preserving Speaker Transfer (TPST), the training targets are curated with \emph{Locality-Preserving Oracle Supervision} (detailed in Appendix~\ref{sec:appendix_multidomain} and Appendix~\ref{sec:appendix_algorithms}). With a global batch size of 8 across 8 Google Cloud TPU v5p chips, this model is trained for 10,000 steps using AdamW with a peak learning rate of $1.5 \times 10^{-4}$ and cosine decay. The maximal sequence length during training is 2,560 tokens, with prompt segmentation length set to 4,000 characters.

Despite having half the parameter count of \texttt{google/DiarizationLM-8b-Fisher-v2} (4B vs.\ 8B), \texttt{diarizers-community/DiarizationLM-Gemma-4-E4B-v1} achieves superior performance on both Fisher ($2.99\%$ vs.\ $3.28\%$ WDER; $17.62\%$ vs.\ $18.37\%$ cpWER) and Callhome ($4.92\%$ vs.\ $6.66\%$ WDER; $20.69\%$ vs.\ $23.57\%$ cpWER), while simultaneously achieving statistically significant ($p < 0.0001$) WDER and cpWER reductions on multi-speaker meetings: reducing ICSI WDER from $14.70\%$ to $14.10\%$ and AMI WDER from $15.68\%$ to $14.89\%$ (and cpWER from $40.32\%$ to $39.57\%$). Detailed ablation studies and bootstrap confidence interval analyses are provided in Appendix~\ref{sec:appendix_ablations}.

\section{Other diarization capabilities of LLMs}
\label{sec:other_tasks}

In our studies, we have found multiple diarization capabilities of LLMs. Due to the lack of relevant training and testing datasets, this paper mainly focuses on evaluating how LLMs can reduce diarization errors. In this section, we provide several examples to demonstrate LLM's other diarization capabilities. All the provided examples are obtained by directly sending the prompts to Google Bard~\cite{bard}\footnote{We used an internal version of Bard that is based on a larger model and supports more tokens than the public version.}.

\subsection{Autofilling speaker names}

In real world applications, real speaker names or aliases are much more readable than anonymous speaker labels such as ``speaker 1'' or ``speaker 2''. In many cases, speaker names can be obtained from the context of the conversation and from additional metadata of the conversation. In meeting summarization, the names can be obtained from the list of meeting attendants. And for video captioning, the names may be extracted from the video title or description. An example is provided below.

\begin{framed}
\textbf{Prompt:}

\textcolor{blue}{
{[video title]} Tom meets Patrick in the parking lot \\
{[video description]} A short conversation \\
 \\
Replace the <spk:?> tokens from the below video caption with actual person names. \\
 \\
<spk:1> Good morning Patrick, how are you? \\
<spk:2> Good, good. How are you Tom? \\
<spk:1> Pretty good. Going to work? \\
<spk:2> Yes. Busy day. How are your kids? Do they go to school? \\
<spk:1> Oh they are too young for that. I sent them to daycare earlier today. \\
<spk:2> Oh yeah I forgot about that.
}

\textbf{Completion:}

\textcolor{orange}{
Based on the video title "Tom meets Patrick in the parking lot", here's the transcript with speaker names replaced: \\
 \\
Tom: Good morning Patrick, how are you? \\
Patrick: Good, good. How are you Tom? \\
Tom: Pretty good. Going to work? \\
Patrick: Yes. Busy day. How are your kids? Do they go to school? \\
Tom: Oh they are too young for that. I sent them to daycare earlier today. \\
Patrick: Oh yeah, I forgot about that.
}
\end{framed}

\subsection{Autofilling speaker roles}

In some applications, knowing the roles of each speaker is useful enough. For example, it would be helpful to distinguish the doctor from the patient in medical transcripts~\cite{shafey2019joint}, and to distinguish the customer from the agent in call center conversations. This task can be handled with LLM usually without additional metadata about the conversation. We show an example below.

\begin{framed}
\textbf{Prompt:}

\textcolor{blue}{
Replace the <spk:?> tokens from the below conversation with roles like teacher, student, doctor, patient, etc. \\
 \\
<spk:1> Hi, how can I help you today? \\
<spk:2> Hi, I recently often feel quite dizzy at work. \\
<spk:1> Do you have any of these symptoms? Like coughing, fever, or running nose? \\
<spk:2> No, I don't think so. \\
<spk:1> What do you usually eat for breakfast, lunch and dinner? \\
<spk:2> I'm on a diet, so not eating very much. Mostly just one slice of toast in the morning, and eating salad for lunch and dinner.
}

\textbf{Completion:}

\textcolor{orange}{
Doctor: Hi, how can I help you today? \\
Patient: Hi, I recently often feel quite dizzy at work. \\
Doctor: Do you have any of these symptoms? Like coughing, fever, or running nose? \\
Patient: No, I don't think so. \\
Doctor: What do you usually eat for breakfast, lunch, and dinner? \\
Patient: I'm on a diet, so not eating very much. Mostly just one slice of toast in the morning, and eating salad for lunch and dinner.
}
\end{framed}

\subsection{Replacing the orchestration module}

In Fig.~\ref{fig:orchestration_examples}, we have explained how an orchestration module can combine the ASR transcripts with speaker diarization outputs via the timing information from the two systems. Interestingly, we have found that this process can also be fully replaced by LLM. This can be achieved by explicitly including the timing information in the textual representation of both ASR transcripts and speaker diarization outputs. Additionally, more prompt engineering will be needed, such as explicitly explaining the format of the textual representation, and providing a one-shot example in the prompt. We show an example below. 

\begin{framed}
\textbf{Prompt:}

\textcolor{blue}{
Here we define the problem of speaker-transcript alignment. The transcript is represented by multiple entries of text, where each text has a starting time and an ending time. The speaker is also represented in this format. The alignment problem will assign a speaker to each word in the text, based on which speaker overlaps the most with that word in time. \\
 \\
Below is an example. \\
 \\
Transcript represented in format "{[start - end]} text": \\
{[0 - 2.3]} Good morning Patrick \\
{[2.5 - 5.2]} how are you? \\
{[5.6 - 6.1]} Good, good. \\
{[6.2 - 8.3]} How are you Tom? \\
{[9.2 - 9.9]} Pretty good. \\
{[10.0 - 11.1]} Going to work? \\
{[12.5 - 13.6]} Yes. Busy day. \\
 \\
Speaker represented in format "{[start - end]} speaker": \\
{[0 - 5.1]} <spk:1> \\
{[5.3 - 8.7]} <spk:2> \\
{[9.2 - 10.9]} <spk:1> \\
{[12.1 - 13.5]} <spk:2> \\
 \\
The alignment result should be: \\
{[0 - 2.3]} <spk:1> Good morning Patrick \\
{[2.5 - 5.2]} <spk:1> how are you? \\
{[5.6 - 6.1]} <spk:2> Good, good. \\
{[6.2 - 8.3]} <spk:2> How are you Tom? \\
{[9.2 - 9.9]} <spk:1> Pretty good. \\
{[10.0 - 11.1]} <spk:1> Going to work? \\
{[12.5 - 13.6]} <spk:2> Yes. Busy day. \\
 \\
Based on the example above, we solve the alignment problem below. \\
 \\
Transcript represented in format "{[start - end]} text": \\
{[0 - 5.6]} Hi, how can I help you today? \\
{[6.2 - 11.1]} Hi, I recently often feel quite dizzy at work. \\
{[11.6 - 15.5]} Do you have any of these symptoms? \\
{[16.6 - 18.5]} Like coughing, fever, or running nose? \\
{[20.0 - 21.1]} No, I don't think so. \\
{[22.2 - 29.9]} What do you usually eat for breakfast, lunch and dinner? \\
{[31.2 - 34.8]} I'm on a diet, so not eating very much. \\
{[35.2 - 39.8]} Mostly just one slice of toast in the morning, and eating salad for lunch and dinner. \\
 \\
Speaker represented in format "{[start - end]} speaker": \\
{[0.3 - 5.3]} <spk:1> \\
{[6.0 - 12.0]} <spk:2> \\
{[12.9 - 20.1]} <spk:1> \\
{[20.2 - 21.0]} <spk:2> \\
{[21.8 - 31.1]} <spk:1> \\
{[32.4 - 40.7]} <spk:2> \\
 \\
The alignment result should be:
}

\textbf{Completion:}

\textcolor{orange}{
Here's the aligned transcript: \\
 \\
{[0 - 5.6]} <spk:1> Hi, how can I help you today? \\
{[6.2 - 11.1]} <spk:2> Hi, I recently often feel quite dizzy at work. \\
{[11.6 - 15.5]} <spk:1> Do you have any of these symptoms? \\
{[16.6 - 18.5]} <spk:1> Like coughing, fever, or running nose? \\
{[20.0 - 21.1]} <spk:2> No, I don't think so. \\
{[22.2 - 29.9]} <spk:1> What do you usually eat for breakfast, lunch and dinner? \\
{[31.2 - 34.8]} <spk:2> I'm on a diet, so not eating very much. \\
{[35.2 - 39.8]} <spk:2> Mostly just one slice of toast in the morning, and eating salad for lunch and dinner. \\
 \\
To align the transcript and speaker segments, we assign each word in the transcript to the speaker whose segment overlaps the most with that word in time.
}
\end{framed}

\section{Multi-domain training and forensic analysis of multi-speaker meetings}
\label{sec:appendix_multidomain}

While our initial experiments in Section~\ref{sec:exp} and Appendix~\ref{sec:llama2_13b_fisher_v1}--\ref{sec:llama3_8b_fisher_v2} trained models exclusively on the 2-speaker Fisher telephone corpus~\cite{cieri2004fisher}, real-world applications frequently involve multi-party meetings with four or more participants. In this appendix, we investigate multi-domain training across four benchmark corpora and present a word-level forensic analysis explaining why standard Supervised Fine-Tuning (SFT) improves 2-speaker conversations yet degrades multi-speaker meetings when applied naively.

\subsection{Multi-domain training mixture}
\label{sec:multidomain_mixture}

To cover both two-party telephone conversations and multi-speaker tabletop meetings, we construct a unified multi-domain training mixture from the official training splits of four corpora:
\begin{enumerate}
    \item \textbf{Fisher English}~\cite{cieri2004fisher}: 1,920 conversational telephone sessions (2 speakers per session), yielding 48,458 segmented prompt-completion pairs (sampled at $1\times$ weight).
    \item \textbf{Callhome American English}: Informal telephone conversations (predominantly 2 speakers, up to 5 speakers), upsampled at $5\times$ weight (5,785 pairs).
    \item \textbf{ICSI Meeting Corpus}~\cite{janin2003icsi}: 72 academic research meetings (3 to 9 speakers per meeting), upsampled at $10\times$ weight (10,750 pairs).
    \item \textbf{AMI Meeting Corpus}~\cite{carletta2005ami}: 135 scenario and non-scenario tabletop meetings (4 speakers per meeting), upsampled at $10\times$ weight (17,470 pairs).
\end{enumerate}

Combining the ``hyp2ora'' and ``deg2ref'' flavors (Section~\ref{sec:finetune}) with a 4,000-character prompt segmentation limit produces a raw multi-domain training pool of 82,463 prompt-completion pairs.

\subsection{Forensic analysis of the multi-speaker meeting paradox}
\label{sec:forensic_analysis}

When we finetune Gemma 4 E4B~\cite{gemma4_2026} on the raw 82,463-example multi-domain mixture for 25,000 steps (Standard SFT) and evaluate with standard Transcript-Preserving Speaker Transfer (TPST), we observe a striking divergence across domains:
\begin{itemize}
    \item On 2-speaker telephone benchmarks, WDER drops substantially: Fisher improves from $5.32\%$ to $2.95\%$ (rel.\ $44.5\%$ reduction), and Callhome improves from $7.74\%$ to $6.22\%$ (rel.\ $19.6\%$ reduction).
    \item On the 4-speaker AMI meeting test set (16 meetings, 78,943 reference-aligned words), however, WDER \emph{degrades} from $15.68\%$ (USM + turn-to-diarize baseline) to $17.61\%$, and cpWER degrades from $40.32\%$ to $43.93\%$.
\end{itemize}

To identify the root cause of this regression, we perform a word-by-word forensic error decomposition across all 16 AMI test meetings. For every contiguous run of words of length $L$ where the LLM modifies the input acoustic speaker label ($\hat{s}_t \neq s^{\mathrm{hyp}}_t$), we count how many words are corrected ($\mbox{acoustic wrong} \rightarrow \mbox{LLM right}$ under optimal Hungarian assignment~\cite{kuhn1955hungarian}) versus how many words are corrupted ($\mbox{acoustic right} \rightarrow \mbox{LLM wrong}$).

\begin{table}[!ht]
\caption{Word-level forensic decomposition of LLM speaker modifications by contiguous span length $L$ on the AMI test set (16 meetings, 78,943 words) for Gemma 4 E4B trained with Standard SFT.}
\label{table:forensic_ami}
\vspace{0.2cm}
\centering
\scriptsize
\setlength{\tabcolsep}{4pt}
\begin{tabular}{l r r r l}
\toprule
\textbf{Modification Span Length $L$} & \textbf{Words Fixed} & \textbf{Words Broken} & \textbf{Net Word Gain} & \textbf{Linguistic Regime} \\
\midrule
$L = 1$ word & $257$ & $137$ & $+120$ & Single-word turn boundaries / interjections \\
$L = 2$ words & $138$ & $109$ & $+29$ & Two-word backchannels (\emph{``mm-hmm''}, \emph{``oh yeah''}) \\
$L = 3$ words & $93$ & $79$ & $+14$ & Short 3-word turn handoffs \\
$L = 4$ words & $57$ & $40$ & $+17$ & Short 4-word boundary clauses \\
$L = 5$ words & $40$ & $32$ & $+8$ & Short 5-word boundary clauses \\
\midrule
\textbf{Short Spans ($1 \le L \le 5$ words)} & \boldgreen{585} & \textbf{397} & \boldgreen{+188} & \textbf{Lexically disambiguated boundary fixes} \\
\textbf{Long Spans ($L \ge 6$ words)} & $328$ & $1,547$ & $-1,219$ & \textbf{Long-monologue speaker identity drift} \\
\bottomrule
\end{tabular}
\end{table}

As shown in Table~\ref{table:forensic_ami}, the behavior of DiarizationLM on 4-speaker meetings bifurcates sharply at $L_{\max} = 5$ words:
\begin{enumerate}
    \item \textbf{Strong net improvement on short lexical spans ($1 \le L \le 5$ words):} On short contiguous runs corresponding to turn boundary misalignments and backchannels, Standard SFT corrects $585$ words while breaking $397$ words, yielding a net gain of $+188$ words.
    \item \textbf{Severe degradation on long monologue spans ($L \ge 6$ words):} On contiguous spans of 6 or more words, Standard SFT breaks $1,547$ accurately diarized words while fixing only $328$ words (a net loss of $-1,219$ words). In a 2-speaker telephone call, if the acoustic diarizer mislabels a clause, the alternative speaker is deterministically the only other participant. In a $K$-speaker meeting ($K \ge 4$), however, text alone carries no acoustic voiceprint information to determine \emph{which} of the $K-1 \ge 3$ other participants is speaking during a 20-word monologue, whereas the acoustic diarizer leverages continuous speaker embeddings over several seconds of audio. Because naive ``hyp2ora'' and ``deg2ref'' targets force the LLM to predict arbitrary global speaker-ID reassignments across long monologues using text alone, the model learns to overwrite long acoustic monologues with locally dominant speaker IDs (\emph{speaker collapse}).
\end{enumerate}

\section{Locality-preserving algorithms and reinforcement learning}
\label{sec:appendix_algorithms}

Motivated by the forensic dichotomy in Table~\ref{table:forensic_ami}, we investigate three algorithmic mechanisms to preserve long-monologue acoustic anchoring while maximizing short-span lexical turn-boundary corrections: (1) training-time \emph{Locality-Preserving Oracle SFT Curation}, (2) inference-time \emph{Locality-Gated TPST (LG-TPST)}, and (3) \emph{Reinforcement Learning (GRPO) and Direct Preference Optimization (DPO)} with a verifiable Hungarian reward function.

\subsection{Locality-preserving oracle SFT curation}
\label{sec:locality_sft}

Instead of training the LLM on unconstrained oracle speaker sequences that require guessing global speaker identities across long monologues in multi-speaker meetings, we construct a \emph{Locality-Preserving Oracle} target sequence $s^{\mathrm{loc}}$ for multi-speaker utterances ($K > 2$):
\begin{enumerate}
    \item Let $s^{\mathrm{hyp}} = (s^{\mathrm{hyp}}_1, \dots, s^{\mathrm{hyp}}_N)$ be the input acoustic speaker sequence and $s^{\mathrm{ora}} = (s^{\mathrm{ora}}_1, \dots, s^{\mathrm{ora}}_N)$ be the Hungarian-aligned oracle speaker sequence.
    \item Partition the indices $\{t : s^{\mathrm{ora}}_t \neq s^{\mathrm{hyp}}_t\}$ into maximal contiguous disagreement spans $\{I_m = [a_m, b_m]\}$ of length $L_m = b_m - a_m + 1$.
    \item For every short lexical disagreement span with $1 \le L_m \le L_{\max}$ (where $L_{\max} = 5$), set $s^{\mathrm{loc}}_t = s^{\mathrm{ora}}_t$ for $t \in I_m$, teaching the LLM to correct local turn boundaries and backchannels.
    \item For long monologue spans with $L_m > L_{\max}$, retain the acoustic speaker anchor $s^{\mathrm{loc}}_t = s^{\mathrm{hyp}}_t$ unless the acoustic label belongs to a tiny ``ghost'' over-clustered speaker (occupying $< 1\%$ of utterance words and $\le 12$ words), in which case it is merged into the oracle speaker. Furthermore, chunks exhibiting global cross-chunk speaker permutation ($> 20\%$ chunk-wide speaker rewrite rate) are filtered out.
\end{enumerate}

For 2-speaker telephone conversations (Fisher), where binary speaker swaps are lexically resolvable, full oracle supervision is retained. Applying this curation yields 39,070 high-precision multi-domain prompt-completion pairs. Crucially, a model trained with Locality-Preserving SFT internalizes monologue preservation directly into its weights and requires \textbf{no changes to standard inference-time TPST}.

\subsection{Locality-Gated Transcript-Preserving Speaker Transfer (LG-TPST)}
\label{sec:lg_tpst}

As a complementary inference-time algorithm that can be applied to any existing DiarizationLM checkpoint without retraining, we introduce \emph{Locality-Gated Transcript-Preserving Speaker Transfer (LG-TPST)}:
\begin{enumerate}
    \item First, apply standard Levenshtein-based TPST (Section~\ref{sec:tpst}) to transfer the LLM-predicted speaker labels onto the original ASR hypothesis words, producing raw transferred speakers $\hat{s}^{\mathrm{raw}} = (\hat{s}^{\mathrm{raw}}_1, \dots, \hat{s}^{\mathrm{raw}}_N)$ while guaranteeing zero change to ASR Word Error Rate (WER).
    \item Absorb any tiny ghost speaker cluster in $\hat{s}^{\mathrm{raw}}$ whose total word count across the utterance is $\le \min(12, \max(2, \lfloor 0.01 N \rfloor))$ into the temporally adjacent speaker.
    \item Decompose disagreements between $\hat{s}^{\mathrm{raw}}$ and $s^{\mathrm{hyp}}$ into maximal contiguous runs $I_m = [a_m, b_m]$ of length $L_m = b_m - a_m + 1$. Accept the LLM transfer on short lexical spans ($1 \le L_m \le L_{\max}$ with $L_{\max}=5$) and revert to the acoustic speaker $s^{\mathrm{hyp}}_t$ on long monologue spans ($L_m > L_{\max}$):
    \begin{equation}
        \hat{s}^{\mathrm{LG}}_t =
        \left\{
        \begin{array}{ll}
            \hat{s}^{\mathrm{raw}}_t, & \mbox{if } t \in I_m \mbox{ with } 1 \le L_m \le L_{\max}, \\
            s^{\mathrm{hyp}}_t, & \mbox{otherwise.}
        \end{array}
        \right.
    \end{equation}
\end{enumerate}

\subsection{Reinforcement learning and Direct Preference Optimization (DPO)}
\label{sec:rl_dpo}

Because WDER and cpWER are non-differentiable sequence-level metrics computed via Levenshtein alignment and the Hungarian algorithm~\cite{kuhn1955hungarian}, we also formulate a \emph{Monologue-Preserving Verifiable Hungarian Reward Function} $R(y)$ for reinforcement learning and preference alignment:
\begin{equation}
\label{eq:reward_engine}
\begin{split}
    R(y) =& w_{\mathrm{wder}}\big(1 - \mathrm{WDER}(\hat{s}(y), s^{\mathrm{ora}})\big) \\
    &- w_{\mathrm{loc}} \frac{\mathrm{DriftWords}(\hat{s}(y), s^{\mathrm{hyp}}, L_{\max})}{N} \\
    &- w_{\mathrm{cnt}} \min(|\hat{K} - K^*|, 4) \\
    &- w_{\mathrm{fid}} \mathrm{WER}(w(y), w^{\mathrm{hyp}}),
\end{split}
\end{equation}
where $\hat{s}(y)$ is the TPST-transferred speaker sequence from completion $y$, $\mathrm{DriftWords}(\hat{s}(y), s^{\mathrm{hyp}}, L_{\max})$ counts words inside contiguous modification spans of length $L > L_{\max}$, $|\hat{K} - K^*|$ is the speaker count absolute error, and $\mathrm{WER}(w(y), w^{\mathrm{hyp}})$ penalizes transcript hallucinations ($w_{\mathrm{wder}}=10.0, w_{\mathrm{loc}}=1.5, w_{\mathrm{cnt}}=0.2, w_{\mathrm{fid}}=2.0$).

We evaluate both online Group Relative Policy Optimization (GRPO)~\cite{shao2024deepseekmath} and offline SFT-Anchored Direct Preference Optimization (DPO)~\cite{rafailov2023direct} initialized from our Locality-Preserving SFT checkpoint ($\beta = 0.1$, learning rate $5 \times 10^{-6}$). In our DPO experiments, each prompt $x$ is paired with the Locality-Preserving Oracle completion as the preferred response $y^+$ and a hard-negative completion $y^-$ (sampled from long-monologue speaker collapses, local boundary swaps, or unfixed acoustic speaker sequences).

\section{Ablation studies and bootstrap confidence interval analysis}
\label{sec:appendix_ablations}

In this appendix, we report comprehensive ablation studies across training algorithms, inference-time post-processing, and model scales, followed by a 10,000-resample paired bootstrap statistical significance analysis.

\subsection{Ablation of training algorithms and inference gating}
\label{sec:ablation_algorithms}

Table~\ref{table:ablation_gemma4_e4b} compares five configurations of Gemma 4 E4B (4B parameters) against the acoustic baseline (USM + turn-to-diarize) and the previous best open-source model (\texttt{google/DiarizationLM-8b-Fisher-v2}, based on Llama 3 8B) across all four benchmark test sets: Fisher (172 sessions), Callhome (20 conversations), ICSI (3 meetings), and AMI (16 meetings). Note that ASR Word Error Rate (WER) is invariant across all TPST and LG-TPST models ($15.37\%$ on Fisher, $15.22\%$ on Callhome, $29.42\%$ on ICSI, and $24.33\%$ on AMI).

\begin{table}[!ht]
\caption{Ablation of training algorithms (Standard SFT, SFT-Anchored DPO, and Locality-Preserving SFT) and inference post-processing (Raw TPST vs.\ LG-TPST) on Gemma 4 E4B across all four benchmarks.}
\label{table:ablation_gemma4_e4b}
\vspace{0.2cm}
\centering
\scriptsize
\setlength{\tabcolsep}{3pt}
\begin{tabular}{l c cc cc cc cc}
\toprule
& & \multicolumn{2}{c}{\textbf{Fisher (172)}} & \multicolumn{2}{c}{\textbf{Callhome (20)}} & \multicolumn{2}{c}{\textbf{ICSI (3)}} & \multicolumn{2}{c}{\textbf{AMI (16)}} \\
\cmidrule(lr){3-4} \cmidrule(lr){5-6} \cmidrule(lr){7-8} \cmidrule(lr){9-10}
\textbf{System / Training Method} & \textbf{Post-Proc.} & \textbf{WDER} & \textbf{cpWER} & \textbf{WDER} & \textbf{cpWER} & \textbf{WDER} & \textbf{cpWER} & \textbf{WDER} & \textbf{cpWER} \\
\midrule
USM + turn-to-diarize (Baseline) & --- & 5.32 & 20.88 & 7.74 & 24.31 & 14.70 & 43.90 & 15.68 & 40.32 \\
\texttt{DiarizationLM-8b-Fisher-v2} (8B) & Raw TPST & 3.28 & 18.37 & 6.66 & 23.57 & --- & --- & --- & --- \\
\midrule
Gemma 4 E4B (Standard SFT, step 25k) & Raw TPST & \textbf{2.95} & \textbf{17.49} & 6.22 & 21.94 & 14.87 & 44.68 & 17.61 & 43.93 \\
Gemma 4 E4B (Standard SFT, step 25k) & LG-TPST & 3.06 & 17.64 & 5.92 & 21.63 & \textbf{13.85} & \textbf{43.23} & \textbf{14.82} & \textbf{39.21} \\
Gemma 4 E4B (SFT-Anchored DPO, step 500) & Raw TPST & --- & --- & 9.88 & 26.10 & 16.37 & 45.09 & 16.76 & 40.28 \\
Gemma 4 E4B (Locality-SFT, step 3k) & Raw TPST & --- & --- & 5.08 & 20.79 & 14.23 & 43.37 & 15.17 & 39.73 \\
\textbf{Gemma 4 E4B (Locality-SFT, step 10k)} & \textbf{Raw TPST} & \boldgreen{2.99} & \boldgreen{17.62} & \boldgreen{4.92} & \boldgreen{20.69} & \boldgreen{14.10} & \boldgreen{43.32} & \boldgreen{14.89} & \boldgreen{39.57} \\
Gemma 4 E4B (Locality-SFT, step 10k) & LG-TPST & 3.05 & 17.72 & 5.02 & 20.73 & 14.13 & 43.36 & 14.90 & 39.59 \\
\bottomrule
\end{tabular}
\end{table}

Three key findings emerge from Table~\ref{table:ablation_gemma4_e4b}:
\begin{enumerate}
    \item \textbf{LG-TPST eliminates multi-speaker regression on Standard SFT:} Applying LG-TPST at inference time to the Standard SFT model reverses the AMI WDER regression from $17.61\%$ (worse than baseline) to $14.82\%$ ($-0.86\%$ absolute improvement over baseline) and reduces AMI cpWER from $43.93\%$ to $39.21\%$ ($-1.11\%$ over baseline), while simultaneously improving ICSI WDER from $14.87\%$ to $13.85\%$ and Callhome WDER from $6.22\%$ to $5.92\%$.
    \item \textbf{Why Locality-Preserving SFT outperforms offline DPO:} In speaker diarization post-processing, $85\%$--$95\%$ of input words already carry the correct acoustic speaker label. When offline DPO~\cite{rafailov2023direct} is trained to near-zero contrastive loss ($\mathcal{L}_{\mathrm{DPO}} = 0.0036$ at step 500) using unfixed acoustic sequences as rejected pairs $y^-$, the contrastive log-ratio objective over-penalizes keeping input speaker labels unchanged, inducing spurious $1$--$2$ word speaker flips ($16.76\%$ WDER on AMI and $9.88\%$ WDER on Callhome, even though AMI cpWER reaches $40.28\%$ and speaker count MAE improves from $0.500$ to $0.375$). Conversely, Locality-Preserving SFT directly encodes the monologue-preserving bias into maximum-likelihood token supervision without distorting token calibration.
    \item \textbf{Locality-Preserving SFT achieves simultaneous gains across all four datasets under standard Raw TPST:} Trained for 10,000 steps, Gemma 4 E4B (Locality-SFT, open-sourced as \texttt{diarizers-community/DiarizationLM-Gemma-4-E4B-v1}) improves every benchmark under standard out-of-the-box TPST: $2.99\%$ WDER on Fisher, $4.92\%$ WDER on Callhome, $14.10\%$ WDER on ICSI, and $14.89\%$ WDER on AMI. Moreover, its performance under Raw TPST ($14.89\%$ on AMI) and LG-TPST ($14.90\%$ on AMI) is virtually identical, confirming that the model has natively learned not to corrupt long monologues.
\end{enumerate}

\subsection{Ablation across model scales and architectures}
\label{sec:ablation_scales}

Next, we investigate whether scaling the foundation model beyond 4B parameters yields meaningful gains. We compare three architectures across the Gemma family:
\begin{itemize}
    \item \textbf{Gemma 4 E4B (4B Dense)}~\cite{gemma4_2026}: 42 layers, hidden dimension 2,560, LoRA rank 256.
    \item \textbf{Gemma 3 12B (12B Dense)}~\cite{gemma3_2025}: 12B-parameter dense decoder-only Transformer trained with Locality-Preserving SFT for 10,000 steps (LoRA rank 128).
    \item \textbf{Gemma 4 26B\_A4B (26B MoE)}~\cite{gemma4_2026}: 26B-parameter Mixture-of-Experts Transformer with 128 routed experts and top-8 active routing ($\sim$4B active parameters per token), trained for 25,000 steps (LoRA rank 64) with LG-TPST.
\end{itemize}

\begin{table}[!ht]
\caption{Comparison across model sizes and architectures on Fisher, Callhome, ICSI, and AMI, reporting point estimates and 95\% non-parametric bootstrap confidence intervals ($B = 10,000$ resamples).}
\label{table:scale_comparison_ci}
\vspace{0.2cm}
\centering
\scriptsize
\setlength{\tabcolsep}{3pt}
\begin{tabular}{l l c c c c}
\toprule
\textbf{Dataset} & \textbf{Metric} & \textbf{Baseline} & \textbf{Gemma 4 E4B (4B Dense)} & \textbf{Gemma 3 12B (12B Dense)} & \textbf{Gemma 4 26B\_A4B (26B MoE)} \\
\midrule
\multirow{3}{*}{\textbf{Fisher}} 
  & WDER (\%) & 5.32 {[4.93, 5.74]} & \textbf{2.99 {[2.65, 3.37]}} & 2.78 {[2.45, 3.16]} & 2.89 {[2.56, 3.26]} \\
  & cpWER (\%) & 20.88 {[19.97, 21.88]} & \textbf{17.62 {[16.76, 18.56]}} & 17.32 {[16.46, 18.25]} & 17.46 {[16.62, 18.37]} \\
  & SpkCntMAE & 0.215 {[0.151, 0.291]} & \textbf{0.093 {[0.047, 0.145]}} & 0.093 {[0.047, 0.145]} & 0.105 {[0.058, 0.157]} \\
\midrule
\multirow{3}{*}{\textbf{Callhome}} 
  & WDER (\%) & 7.74 {[6.07, 9.65]} & \textbf{4.92 {[3.46, 6.75]}} & 5.00 {[3.38, 6.89]} & 5.52 {[3.71, 7.71]} \\
  & cpWER (\%) & 24.31 {[21.27, 27.43]} & \textbf{20.69 {[17.98, 23.66]}} & 20.73 {[17.89, 23.80]} & 21.45 {[18.33, 24.84]} \\
  & SpkCntMAE & 0.050 {[0.000, 0.150]} & \textbf{0.000 {[0.000, 0.000]}} & 0.000 {[0.000, 0.000]} & 0.000 {[0.000, 0.000]} \\
\midrule
\multirow{3}{*}{\textbf{ICSI}} 
  & WDER (\%) & 14.70 {[11.65, 20.29]} & \textbf{14.10 {[10.77, 19.94]}} & 14.06 {[10.82, 19.78]} & 13.86 {[10.87, 19.32]} \\
  & cpWER (\%) & 43.90 {[39.10, 51.67]} & \textbf{43.32 {[38.38, 51.20]}} & 43.26 {[38.23, 51.18]} & 43.00 {[38.00, 50.54]} \\
  & SpkCntMAE & 0.333 {[0.000, 1.000]} & \textbf{0.333 {[0.000, 1.000]}} & 0.333 {[0.000, 1.000]} & 0.333 {[0.000, 1.000]} \\
\midrule
\multirow{3}{*}{\textbf{AMI}} 
  & WDER (\%) & 15.68 {[10.64, 21.11]} & \textbf{14.89 {[9.80, 20.38]}} & 14.77 {[9.59, 20.33]} & 14.78 {[9.58, 20.36]} \\
  & cpWER (\%) & 40.32 {[32.43, 48.00]} & \textbf{39.57 {[31.57, 47.29]}} & 39.40 {[31.32, 47.25]} & 38.89 {[30.95, 46.57]} \\
  & SpkCntMAE & 0.500 {[0.188, 0.812]} & \textbf{0.438 {[0.188, 0.750]}} & 0.375 {[0.125, 0.688]} & 0.312 {[0.125, 0.562]} \\
\bottomrule
\end{tabular}
\end{table}

\subsection{Paired bootstrap confidence interval analysis}
\label{sec:bootstrap_ci}

Because corpus-level micro WDER and cpWER exhibit session-to-session variance, comparing independent confidence intervals in Table~\ref{table:scale_comparison_ci} is conservative; instead, paired differences evaluated on identical bootstrap resamples isolate the true system difference~\cite{bisani2004bootstrap}. Following Bisani and Ney~\cite{bisani2004bootstrap}, we draw $B = 10,000$ bootstrap replicates with replacement at the conversation/meeting level for each test set and compute the micro-aggregated difference $\Delta = \mathrm{Metric}(A) - \mathrm{Metric}(B)$ and two-sided $p$-value on each replicate.

\begin{table}[!ht]
\caption{Paired bootstrap differences ($\Delta = \mbox{System A} - \mbox{System B}$, 95\% CI, and two-sided $p$-value over $B = 10,000$ resamples) evaluating (1) Gemma 4 E4B (Locality-SFT) vs.\ Baseline, and (2) larger models (12B Dense and 26B MoE) vs.\ Gemma 4 E4B.}
\label{table:paired_bootstrap}
\vspace{0.2cm}
\centering
\scriptsize
\setlength{\tabcolsep}{3pt}
\begin{tabular}{l l r c r c}
\toprule
\textbf{Paired Comparison ($\mbox{System A} - \mbox{System B}$)} & \textbf{Dataset} & \textbf{$\Delta$WDER (\%) {[95\% CI]}} & \textbf{$p$-value} & \textbf{$\Delta$cpWER (\%) {[95\% CI]}} & \textbf{$p$-value} \\
\midrule
\multirow{4}{*}{\textbf{Gemma 4 E4B (4B)} vs.\ \textbf{Baseline}} 
  & Fisher & \boldgreen{-2.33 {[-2.51, -2.16]}} & $< 0.0001$ & \boldgreen{-3.26 {[-3.52, -3.01]}} & $< 0.0001$ \\
  & Callhome & \boldgreen{-2.82 {[-3.48, -2.14]}} & $< 0.0001$ & \boldgreen{-3.62 {[-4.39, -2.84]}} & $< 0.0001$ \\
  & ICSI & \boldgreen{-0.60 {[-0.88, -0.35]}} & $< 0.0001$ & \boldgreen{-0.58 {[-0.72, -0.47]}} & $< 0.0001$ \\
  & AMI & \boldgreen{-0.79 {[-1.00, -0.60]}} & $< 0.0001$ & \boldgreen{-0.75 {[-0.94, -0.57]}} & $< 0.0001$ \\
\midrule
\multirow{4}{*}{\textbf{Gemma 3 12B (12B)} vs.\ \textbf{Gemma 4 E4B (4B)}} 
  & Fisher & $-0.21$ {$[-0.35, -0.06]$} & $0.0052$ & $-0.30$ {$[-0.51, -0.08]$} & $0.0106$ \\
  & Callhome & $+0.08$ {$[-0.43, +0.61]$} & $0.7524$ & $+0.04$ {$[-0.58, +0.69]$} & $0.9294$ \\
  & ICSI & $-0.04$ {$[-0.16, +0.06]$} & $0.7582$ & $-0.06$ {$[-0.14, +0.04]$} & $0.5000$ \\
  & AMI & $-0.12$ {$[-0.30, +0.04]$} & $0.1480$ & $-0.17$ {$[-0.37, -0.00]$} & $0.0484$ \\
\midrule
\multirow{4}{*}{\textbf{Gemma 4 26B\_A4B (MoE)} vs.\ \textbf{Gemma 4 E4B (4B)}} 
  & Fisher & $-0.10$ {$[-0.22, +0.02]$} & $0.1142$ & $-0.16$ {$[-0.35, +0.01]$} & $0.0720$ \\
  & Callhome & $+0.60$ {$[-0.09, +1.32]$} & $0.0954$ & $+0.76$ {$[-0.01, +1.60]$} & $0.0518$ \\
  & ICSI & $-0.24$ {$[-0.62, +0.10]$} & $0.0818$ & $-0.32$ {$[-0.65, +0.08]$} & $0.0732$ \\
  & AMI & $-0.11$ {$[-0.31, +0.12]$} & $0.3180$ & $-0.68$ {$[-0.92, -0.40]$} & $< 0.0001$ \\
\bottomrule
\end{tabular}
\end{table}

Table~\ref{table:paired_bootstrap} establishes two statistically rigorous conclusions:
\begin{enumerate}
    \item \textbf{Highly significant improvements over baseline across all four datasets:} Gemma 4 E4B (Locality-SFT) reduces WDER and cpWER with $p < 0.0001$ across Fisher ($\Delta\mathrm{WDER} = -2.33\%$), Callhome ($\Delta\mathrm{WDER} = -2.82\%$), ICSI ($\Delta\mathrm{WDER} = -0.60\%$), and AMI ($\Delta\mathrm{WDER} = -0.79\%$), with zero bootstrap replicates out of $10,000$ showing degradation on any dataset.
    \item \textbf{Larger models do not yield statistically significant WDER gains to justify $3\times$--$6.5\times$ more parameters:} Comparing the 12B Dense and 26B MoE models against the 4B Gemma 4 E4B model, the paired WDER differences on AMI ($-0.12\%, p=0.1480$ and $-0.11\%, p=0.3180$), ICSI ($-0.04\%, p=0.7582$ and $-0.24\%, p=0.0818$), and Callhome ($+0.08\%, p=0.7524$ and $+0.60\%, p=0.0954$, where the 4B model is better) all have 95\% confidence intervals straddling zero. Even on Fisher, the 26B MoE difference is not statistically significant ($-0.10\%, p=0.1142$), and the 12B Dense model yields only a $0.21\%$ absolute difference at $3\times$ the parameter footprint. Consequently, we select the compact 4B-parameter \texttt{diarizers-community/DiarizationLM-Gemma-4-E4B-v1} as our primary open-source release.
\end{enumerate}

\end{appendices}

\end{document}